\documentclass[journal]{IEEEtran}
\IEEEoverridecommandlockouts
\usepackage{lettrine}
\usepackage{blindtext}
\usepackage{graphicx}
\usepackage{algorithm}
\usepackage{algorithmic}
\usepackage{graphicx}
\usepackage{epstopdf}
\usepackage{subfigure}
\usepackage{amsmath,bm}
\usepackage{amssymb}
\usepackage{amsthm}
\usepackage{cases}
\usepackage{multirow}
\newtheorem{theorem}{\textbf{Theorem}}

\newtheorem{lemma}{Lemma}

\usepackage{extpfeil}
\usepackage[numbers,sort&compress]{natbib}

\usepackage{extarrows}
\usepackage{array}
\usepackage{booktabs}
\allowdisplaybreaks[3]

\title{ Modeling, Analysis, and Optimization of Grant-Free NOMA in Massive MTC via Stochastic Geometry}

\author{\IEEEauthorblockN{Jiaqi Liu, \IEEEmembership{Student Member,~IEEE,}
        Gang Wu, \IEEEmembership{Member,~IEEE,}
        Xiaoxu Zhang, \IEEEmembership{Member,~IEEE,}\\
        Shu Fang, \IEEEmembership{Member,~IEEE,}
		and Shaoqian Li, \IEEEmembership{Fellow,~IEEE}
	\thanks{This paper is supported in part by National Key Research and Development Program under Grant No. 2018YFB1800800. \textit{(Corresponding author: Gang Wu.)}}
	\thanks{ J. Liu, G. Wu, S. Fang, and S. Li are with the National Key Laboratory of Science and Technology on Communications, University of Electronic Science and Technology of China, Chengdu 611731, China (e-mail: ljq\_uestc@163.com; wugang99@uestc.edu.cn; susanfang212@gmail.com; lsq@uestc.edu.cn).}
    \thanks{ X. Zhang is with the School of Information Science and Technology, Southwest Jiaotong University, Chengdu 611756, China (e-mail: xiaoxuzhang@swjtu.edu.cn).}
	}}

\begin{document}
\renewcommand{\figurename}{Fig.}
\maketitle

\begin{abstract}
	
Massive machine-type communications (mMTC) is a crucial scenario to support booming Internet of Things (IoTs) applications. In mMTC, although a large number of devices are registered to an access point (AP), very few of them are active with uplink short packet transmission at the same time, which requires novel design of protocols and receivers to enable efficient data transmission and accurate multi-user detection (MUD). Aiming at this problem, grant-free non-orthogonal multiple access (GF-NOMA) protocol is proposed. In GF-NOMA, active devices can directly transmit their preambles and data symbols altogether within one time frame, without grant from the AP. Compressive sensing (CS)-based receivers are adopted for non-orthogonal preambles (NOP)-based MUD, and successive interference cancellation is exploited to decode the superimposed data signals. In this paper, we model, analyze, and optimize the CS-based GF-MONA mMTC system via stochastic geometry (SG), from an aspect of network deployment. Based on the SG network model, we first analyze the success probability as well as the channel estimation error of the CS-based MUD in the preamble phase and then analyze the average aggregate data rate in the data phase. As IoT applications highly demands low energy consumption, low infrastructure cost, and flexible deployment, we optimize the energy efficiency and AP coverage efficiency of GF-NOMA via numerical methods. The validity of our analysis is verified via Monte Carlo simulations. Simulation results also show that CS-based GF-NOMA with NOP yields better MUD and data rate performances than contention-based GF-NOMA with orthogonal preambles and CS-based grant-free orthogonal multiple access.

\end{abstract}

\begin{IEEEkeywords}
Massive machine-type communications, grant-free, non-orthogonal multiple access, compressed sensing, stochastic geometry.
\end{IEEEkeywords}

\section{Introduction}

\lettrine[lines=2]{M}{ASSIVE} machine type communications (mMTC) is an emerging technology to support the proliferation of Internet of things (IoT) applications by providing a unified interconnection framework as well as facilitating a seamless connectivity of intelligent devices and management platforms \cite{palattella2016internet}. 
In a typical mMTC system, a great number of user devices are registered to an access point (AP), only a very small fraction of them expecting to transmit short data packets to the AP in each time slot \cite{yu2017fundamental}.
IoT applications generally have high demands in low latency, high reliability, and low power assumption. 
These features make mMTC much different from the human-centric communication scenarios dominating the cellular Internet of today and call for novel access schemes and protocols for this potential scenario.

In the long time evolution (LTE) system, resource request and scheduling are needed before uplink data transmission because different users must transmit their data over orthogonally divided radio resources to avoid collision.
The grant-based LTE uplink transmission requires a four-handshake procedure consisting of scheduling request, uplink grant, uplink data transmission, and ACK/NACK transmission,
which has a typical end-to-end latency of 17 ms in total to transmit a data packet of one-frame length \cite{3gpp2016latency}.
Among the 17 ms, only 4 ms are used for data signaling and decoding, whereas 8 ms are used for request transmission and handling.
Obviously, if the mMTC scenario directly adopts the LTE uplink procedure,
then the massive-connectivity and low-latency requirements cannot be satisfied, and the radio resources cannot be fully utilized.
In this regard, grant-free non-orthogonal multiple access (GF-NOMA) has been considered as a promising solution to achieve massive connectivity, low latency, and high spectrum efficiency \cite{wang2017performance}.
With non-orthogonal multiple access (NOMA) technique, the superimposed data from multiple user devices over the same radio resource is still decodable, which enables grant-free transmission \cite{au2014uplink,ding2014performance}.
Then user devices can transmit their data as soon as data packets arrive, and the data can be transmitted together with the preambles in one shot, which can significantly reduce the end-to-end latency of data transmission.
Although the overloading gain of GF-NOMA is at the expense of increased processing complexity of non-linear receivers, the complexity is affordable at the AP side for the uplink-dominated mMTC scenarios.

Recently, among the intensive studies on GF-NOMA, there are two main categories: the contention-based approaches and the contention-free approaches.
Orthogonal preambles (OPs) are used to active user detection (AUD) and channel estimation (CE) by the contention-based approaches,
which are thus known as OP-based GF-NOMA (OP-GF-NOMA) \cite{shirvanimoghaddam2017fundamental,jiang2018analyzing,gharbieh2018spatiotemporal,moussa2019rach,abbas2019novel}.
Since the number of available OPs is upper-bounded by the preamble length to guarantee the orthogonality and the number of user devices in mMTC is generally much larger than the maximum system-supported preamble length,
each user device cannot be pre-assigned with a certain preamble and will randomly choose a preamble from the preamble pool when active.
The advantages of OP-GF-NOMA are that the orthogonality of preambles helps to improve the detection reliability and that the system design can refer to the random access channel (RACH) in the LTE system.
However, OP-GF-NOMA cannot eliminate preamble collision.
When more than one user devices choose a preamble, collision happens.
So that the AP cannot distinguish among the user devices in collision, and then missed detection will happen.
Although the backoff and retransmission procedures are designed to resolve the collisions and thus to improve reliability,
they will inevitably lead to high latency.

In this paper, we focus on the contention-free approaches, which use non-orthogonal preambles (NOPs) to enable preamble overload and support contention-free transmission \cite{liu2018sparse,wang2016dynamic,wei2017approximate,chen2018sparse,cirik2018multi,wei2019message,irtaza2019greedy}.
Exploiting the sporadic nature of device activity in mMTC, compressive sensing (CS)-based sparsity reconstruction algorithms can be adopted to develop efficient AUD and CE algorithms.
Therefore, this kind of GF-NOMA with NOPs is known as CS-based GF-NOMA (CS-GF-NOMA).
It has been proved that as long as the restricted isometry property (RIP) is satisfied among the NOPs,
the CS-based AUD is effective.
So that number of available NOPs in the CS-GF-NOMA system is no longer limited by the preamble length \cite{candes2005decoding}.
Therefore, each user device can be pre-assigned with a unique preamble, and the preamble transmission can be free from collision.
Compared with OP-GF-NOMA, CS-GF-NOMA can further reduce end-to-end latency by avoiding retransmission caused by preamble collision.
Moreover, with NOPs pre-assigned to user devices, the sleeping and activating mechanism of the user devices can be more elastic to reduce the energy comsumption and maintanance cost of real-time IoT applications.
However, the non-orthogonality of preamble sequences brings challenges on the design of reliable AUD algorithms and NOPs \cite{yu2017pilot,chen2018new}.

Developing from the well-studied sparsity reconstruction algorithms in CS field, many efficient AUD algorithms for CS-GF-NOMA have been proposed.
Based on orthogonal matching pursuit (OMP), Wang \textit{et al.} \cite{wang2016dynamic} proposed a low-complexity dynamic AUD algorithm for jointly user activity and data detection in GF-NOMA.
By approximate message passing (AMP) and expectation maximization (EM), Wei \textit{et al.} \cite{wei2017approximate} significantly improve the performance of jointly user activity and data detection in GF-NOMA by exploiting the structured sparsity of user activity and the prior information on transmitted data.
Chen \textit{et al.} \cite{chen2018sparse} proposed an AMP-based algorithm exploiting the statistics of the wireless channel to improve the reliability of AUD in GF-NOMA.
Cirik \textit{et al.} \cite{cirik2018multi} proposed an alternative direction method of multipliers (ADMM)-based AUD algorithm to jointly detect user activity and transmitted data, which exploits the prior information of AUD results in previous time interval to improve the AUD performance in current time interval.
Wei \textit{et al.} \cite{wei2019message} proposed an expectation propagation (EP) algorithm for the joint CE and data decoding of grant-free SCMA.
Irtaza \textit{et al.} \cite{irtaza2019greedy} proposed an enhanced greedy OMP algorithm for joint AUD, CE, and data decoding.
Although a variety of efficient CS-based AUD algorithms have been proposed, most of the aforementioned work only validates the algorithms by simulations.
There lacks theoretical analysis to validate the effectiveness of CS-based AUD for GF-NOMA,
especially for the model and analysis methodology from a network deployment aspect to guide the network-level optimization of GF-NOMA.

Stochastic geometry (SG) \cite{elsawy2017modeling} and queueing theory \cite{alfa2020queueing} are two widely used mathematical tools to model and analyze mMTC systems with randomly deployed user and bursting data transmission.
There has been some early work using SG to model and analyze NOMA networks or grant-free strategies.
For example, Ding \emph{et al.} \cite{ding2014performance} analyze the outage probability and ergodic sum rate of NOMA networks,
Yang \emph{et al.} \cite{yang2017uplink} analyze the ergodic sum rate of sparse code multiple access (SCMA) networks,
and Abbas \emph{et al.} \cite{abbas2019novel} analyze the outage probability and throughput of OP-GF-NOMA networks.
Seo \textit{et al.} \cite{seo2019low} use queueing theory to model and analyze the latency of CS-GF-NOMA with backoff and retransmission.
Recently, directly adopting the LTE RACH protocol into grant-free mMTC and analyzing the performance with spatio-temporal models is intensively investigated.
The spatio-temporal model combines SG and queueing theory together to comprehensively consider the per-device packet arrival rate, the spatial device distribution, the access control, and the backoff and retransmission protocols of the LTE RACH-based grant-free mMTC systems \cite{jiang2018analyzing,gharbieh2018spatiotemporal,moussa2019rach}.
However, little is known about the model and analysis of the CS-GF-NOMA mMTC systems from network deployment aspect, to the best of our knowledge.

In this paper, we use SG to model the CS-GF-NOMA mMTC system and analyze the single-time frame performance determined by network geometric deployment.
As mMTC scenarios generally have strict requirements of low latency,
it is more valuable to analyze the instantaneous performance within a single grant-free time frame than to analyze the steady-state performance with consideration of backoff and retransmission procedures.
Moreover, we optimize the energy efficiency (EE) and the access point coverage efficieny (APCE) of the GF-NOMA network via numerical method, which will be presented with the numerical results in Section \ref{Sec_Numerical}.
The energy consumption is especially important for the system design of IoT because the user devices are required to sustain long battery life for the purpose of lower maintenance cost \cite{wang2019secrecy}.
The APCE concerns the ability of an AP to support massive connectivity.
In multi-cell cases, higher APCE of each AP indicates that fewer APs are needed to be deployed to cover the whole network area.
The contributions of this paper can be summarized as follows:
\begin{itemize}
	\item A novel SG network model is proposed to model the randomly deployed devices and the bursting uplink grant-free transmission in the CS-GF-NOMA mMTC system.
	
	\item Based on the SG network model, the perfect AUD probability, the CE error, and the average aggregate data rate of the CS-GF-NOMA mMTC system are derived with closed-form expressions.
	
	\item The EE and the APCE of the CS-GF-NOMA mMTC system are optimize via numerical methods to meet the low-energy-consumption and low-infrastructure-cost demands of IoT applications.
	
	\item Simulation results are presented to show the validity of our analysis and the advantages of CS-GF-NOMA over OP-GF-NOMA and grant-free orthogonal multiple access (GF-OMA).
\end{itemize}

The remainder of this paper is organized as follows. 
The system model is introduced in Section II.
The perfect AUD probability and CE error of CS-GF-NOMA is mathematically analyzed in Section III.
The aggegrate data rate of CS-GF-NOMA is mathematically analyzed in Section IV.
In Section V, the analytical results of CS-GF-NOMA are validated with simulation results and compared with that of OP-GF-NOMA and GF-OMA, and the EE and APCE of CS-GF-NOMA are numerically optimized.
Section VI concludes this paper.


\section{System Model}

In this section, we present the system model of the CS-GF-NOMA mMTC system.
First, we exploit SG to formulate the network geometry,
which depicts the spatial distribution of the devices in the system.
Then we illustrate the signal model, including the signaling of preambles and data, as well as a path loss channel model with Rayleigh fading.
Finally, we illustrate the energy consumption model.

\subsection{Network Geometry}

\begin{figure}
  \centering
  \includegraphics[scale=0.85]{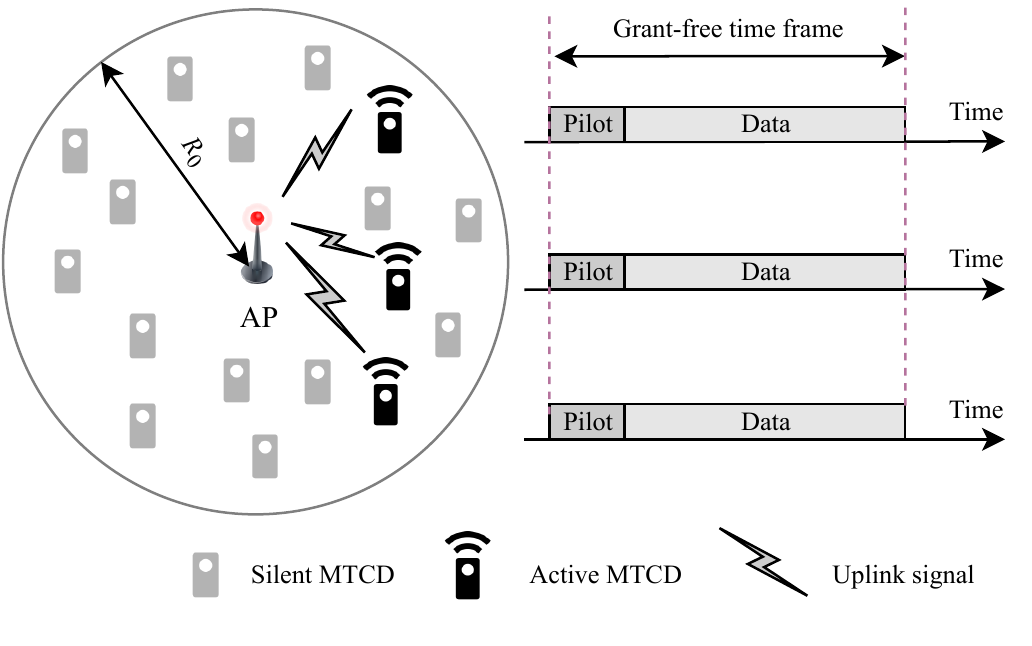}\\
  \caption{System model of an uplink CS-GF-NOMA mMTC system. $K=3$ active MTCDs among $N$ potential MTCDs transmit their data symbols and preambles within a grant-free time frame, while the other MTCDs keep silent.}
  \label{system_model}
\end{figure}

We consider uplink GF-NOMA transmission in an mMTC system shown as Fig. \ref{system_model}, where a signal-antenna AP located at the origin serves $N$ signal-antenna machine-type communication devices (MTCDs).
The MTCDs are uniformly distributed in an annulus of inner radius $D_0$ and outer radius $D_1$.
The cumulative distribution function (CDF) of the distance $r$ between an MTCD and the AP is
\begin{align}
F_r(r) = \frac{\pi r^2 -\pi D_0^2}{\pi D_1^2 -\pi D_0^2}, ~ D_0 \leq r \leq D_1,
\end{align}
and thus the probability density function (PDF) of $r$ is
\begin{align}
f_r(r) = \frac{\mathrm{d}}{\mathrm{d}r}F_r(r)=\frac{2r}{D_1^2-D_0^2}, ~ D_0 \leq r \leq D_1.
\end{align}

When a data packet arrives at an MTCD, this MTCD is activated and then transmit the data packet to the AP in the nearest grant-free time slot.
Denote $\mathcal{P}_\mathrm{ACT}$ as the probability that a data packet arrives at an MTCD within each time frame.
Since the number of MTCDs $N\gg1$ is very large whereas $\mathcal{P}_\text{ACT}$ is very low in the mMTC scenario,
the active MTCDs within each time frame can be approximately modeled as a two-dimensional homogeneous Poisson point process (HPPP) of intensity $\lambda = N\mathcal{P}_\text{ACT}$.

\subsection{Signal Model}

Let $\mathcal{N} = \{1,\cdots,N\}$ denote the set of $N$ potential MTCDs.
We assume that each MTCD $n\in\mathcal{N}$ is assigned with a unique length-$M$ pseudo-random preamble sequence $\boldsymbol{\phi}_n = [\phi_{n,1},\cdots,\phi_{n,M}]^\text{T}$.
The preamble sequences are unified such that $\|\boldsymbol{\phi}_1\|^2=\cdots=\|\boldsymbol{\phi}_N\|^2=1$.
Gathering the pilot sequence for all MTCDs, we can obtain an $M\times N$ pilot matrix
$\boldsymbol{\Phi} = \left[\boldsymbol{\phi}_1, \cdots, \boldsymbol{\phi}_N \right]$.
The preamble sequences are used for both MTCD identification and data symbol spreading.

We assume that the overall grant-free frequency band is divided into $M$ orthogonal sub-channels
and each time frame consists of $L+1$ symbols,
where the first symbol is used for preamble transmission and the following $L$ symbols are used for data transmission.
The $M$ sub-channels are within coherent bandwidth, and the $L+1$ symbols are within coherent time.
Each active MTCD transmits one preamble following by $L$ data symbols within one time frame.
The $l$th data symbol $s_{n,l}$ of the $n$th MTCD is taken from a complex constellation set $\mathcal{X}_n$ and spread with preamble $\boldsymbol{\phi}_n$.
Specifically, when the $n$th MTCD is active, it transmits $\boldsymbol{\phi}_n$ in the preamble symbol and $s_{n,l}\boldsymbol{\phi}_n$ in the $l$th data symbol, over the $M$ sub-channels.

The activity of the $n$th MTCD is represented by a binary parameter $a_n$,
where $a_n=1$ for the active MTCDs and $a_n=0$ for the silent MTCDs.
In the pilot phase, the received signal of the AP on the $M$ sub-channels can be stacked in an $M$-dimensional  complex vector
\begin{align}\label{rx_pilot}
	\mathbf{y}_0 = \sum\limits_{n=1}^{N} a_n h_n \sqrt{P} \boldsymbol{\phi}_n  + \mathbf{w}_0
	  = \mathbf{\Phi} \mathbf{q} + \mathbf{w}_0,
\end{align}
where $h_n$ is the complex channel coefficient of the $n$th MTCD,
$P$ is the transmit power of each MTCD,
$\mathbf{w}_0$ is the noise consisting of i.i.d. complex Gaussian distributed entries following $\mathcal{CN}(0,\sigma^2)$,
$\mathbf{q} = [q_1,\cdots,q_N]^\mathrm{T}$,
and $q_n= a_n h_n \sqrt{P}$ is the joint channel gain, user activity, and transmit power of the $n$th MTCD.

We adopt a standard power-law path-loss model to model the channel,
where the path-loss is inversely proportional to link distance with the path-loss exponent $\alpha$,
and the multi-path fading accords with the Rayleigh fading.
Therefore, the channel power gain of the $n$th MTCD can be expressed as
$|h_n|^2 = \xi_n r_n^\alpha$, where $r_n$ is the distance from the $n$th MTCD to the AP
and $\xi_n\sim\exp(1)$ is a random variable from exponential distribution with unit mean.

In the data phase, the $l$th received data symbol of the AP on those $M$ sub-channels can be stacked in an $M$-dimensional complex vector
\begin{align}\label{rx_data}
\mathbf{y}_l = \sum\limits_{n=1}^{N} a_n h_n \sqrt{P} s_{n,l} \boldsymbol{\phi}_n  + \mathbf{w}_l,
\end{align}
where $\mathbf{w}_l$ is the noise of the $l$th symbol.

The objective of the AP is to identify all the active MTCDs, 
i.e. the MTCDs with $a_n=1$, based on the received preamble signals
and to decode the data symbols based on the received data signals.

\subsection{Energy Consumption Model}

We assume that each MTCD has totally three components of energy consumption: the first one is the static energy consumption in inactive state; the second and the third ones are respectively the dynamic circuit energy consumption and the antenna energy consumption in active state \cite{anamuro2018modeling,shahini2019noma}.

The average static power for an inactive MTCD is $P_\mathrm{S}$ to operate and keep synchronized with the AP.
When an MTCD becomes active within a time frame and communicates to the AP within the nearest time frame, 
the average dynamic circuit power for data handling, signal processing and modulation is $P_\mathrm{D}$ over the two time frames,
and the antenna input power is $P_\mathrm{A}$ within the latter time frame.
The relationship between $P$ and $P_\mathrm{A}$ is $P=\varepsilon P_\mathrm{A}$, where $\varepsilon$ is the antenna efficiency.
Then the long-term average power consumption of each MTCD is
\begin{align}
	 \overline{P}_\mathrm{DEV} = \left(1- \mathcal{P}_\mathrm{ACT}\right) P_\mathrm{S} + 
		\mathcal{P}_\mathrm{ACT} \left( 2 P_\mathrm{D} + P_\mathrm{A} \right).
\end{align}

\section{Performance Analysis for Active User Detection and Channel Estimation}

In this section, we first formulate the AUD and CE of CS-GF-NOMA mMTC systems as a least absolute shrinkage and selection operator (LASSO) problem.
Then based on the threshold analysis of the LASSO problem \cite{wainwright2009information,wainwright2009sharp},
we derive the closed-form expression of the AUD success probability and CE error based on the SG-based network model.

\subsection{The LASSO Problem}

In the preamble phase, the AP detects the activity and estimate the channel coefficient of the active MTCDs, based on the received preamble signal.
For analytic tractability, we formulate the preamble-based joint AUD and CE of CS-GF-NOMA as an LASSO problem, which solves a $\ell_1$-constrained quadratic program given by
\begin{align}\label{lasso}
	\widehat{\mathbf{q}} = \mathop{\mathrm{argmin}}_{\mathbf{q}} \left\lbrace
		\frac1{2M} \|\mathbf{y}_0 -\mathbf{\Phi} \mathbf{q} \|^2_2
		+ \gamma \| \mathbf{q} \|_1 \right\rbrace,
\end{align}
where $\gamma = \sqrt{2 c_1\sigma^2\log N/M}$ is a regularization parameter with constant $c_1\geq2$.

Based on Wainwright's analysis on the performance of LASSO for CS-based sparse signal recovery \cite{wainwright2009information,wainwright2009sharp},
we introduce two important parameters that determines the solution to the LASSO problem (\ref{lasso}), which are respectively the maximum supported sparsity as
\begin{align}\label{K_max}
	K_\text{max} = \left\lfloor \frac{M}{2\log N}\left(c_2 -\frac1{c_1}\right) \right\rfloor
\end{align}
and the minimum detectable amplitude threshold as
\begin{align}\label{q_min}
	\upsilon = c_3 \gamma + 20\sqrt{\sigma^2\log K/M},
\end{align}
with constant $c_2,c_3>0$.
In large system with sporadic device activity, i.e. $K\ll N$, we can simplify (\ref{q_min_succ}) as $\upsilon = c_3 \gamma$.

Let $\mathsf{Supp}(\mathbf{q})$ denotes the support set of $\mathbf{q}$, $K = |\mathsf{Supp}(\mathbf{q})|$, and $q_{\text{min}} = \min_{n\in\mathsf{Supp}(\mathbf{q})} |q_n|$.
In the CS-GF-NOMA mMTC system, $\mathsf{Supp}(\mathbf{q})$, $K$, and $q^2_{\text{min}}$ are respectively the set, the number, and the minimum AP received power of active MTCDs.
Based on the thresholds $K_\text{max}$ and $\upsilon$, there may be three events happening for the solution to the LASSO problem (\ref{lasso}), shown in Fig. \ref{venn_diagram}.
The conditions for perfect AUD and failure AUD are given by Lemma 1, proved in Appendix A.

\begin{figure}
  \centering
  \includegraphics[scale=1]{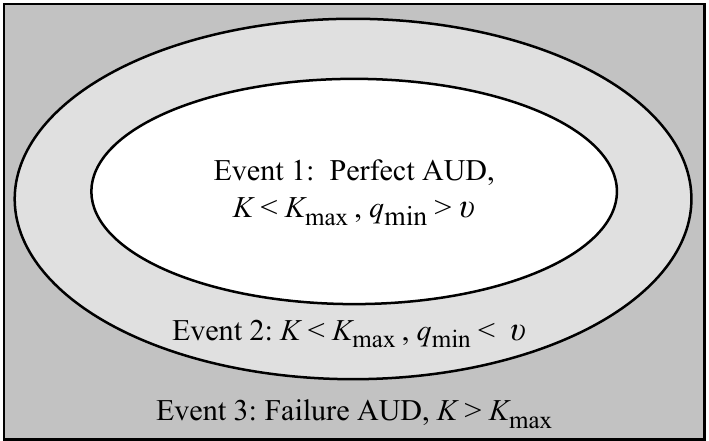}\\
  \caption{Venn diagram showing the events of the LASSO receiver.
  When Event 1 happens, LASSO can exactly detect all the active MTCDs and the error of CE can be very small.
  When Event 2 happens, LASSO can detect a part of the active MTCDs with missed detection and the missed MTCDs influences the accuracy of the CE and the data decoding upon the detected MTCDs.
  When Event 2 happens, the results of LASSO may be a mixture of false detection and missed detection,
  which is out of our consideration.}\label{venn_diagram}
\end{figure}

\begin{lemma}
	When $K$ MTCDs are active in the CS-GF-NOMA mMTC system,
	if the following conditions are satisfied:
	\begin{align}
	    & K \leq K_\mathrm{max},\label{K_max_succ}\\
	    & q_\mathrm{min} > \upsilon,\label{q_min_succ}
	\end{align}
	then the LASSO AUD can perfectly detect all the active MTCDs with probability converging to one.
	If the following condition is satisfied:
	\begin{align}
	    K>K_\text{max},\label{K_max_fail}
	\end{align}
	then the LASSO AUD fails with probability converging to one.
\end{lemma}

According to \citep[Theorems 3]{wainwright2009sharp},
condition (\ref{K_max_succ}) guarantees that the LASSO AUD can successfully reconstruct a subset of $\mathsf{Supp}(\mathbf{q})$ without false detection,
namely $\mathsf{Supp}(\widehat{\mathbf{q}})\subseteq \mathsf{Supp}({\mathbf{q}})$.
Then further with condition (\ref{q_min_succ}) satisfied, 
the LASSO AUD can exactly reconstruct $\mathsf{Supp}(\mathbf{q})$ with neither false detection nor missed detection, namely $\mathsf{Supp}(\widehat{\mathbf{q}})= \mathsf{Supp}({\mathbf{q}})$,
and the LASSO CE is with limited estimation error.
According to \citep[Theorems 4]{wainwright2009sharp},
when condition condition (\ref{K_max_succ}) cannot be satisfied, i.e. $K > K_\mathrm{max}$, 
the result of LASSO AUD is a mixture of false detection and missed detection, which is difficult to analyze.

We can observe from (\ref{K_max}) that $K_\text{max}$ increases with the decrease of $N$ and the increase of $M$.
It implies that in mMTC systems with sporadic device activity,
with a larger preamble length $M$ and a smaller number $N$ of potential MTCDs,
more active MTCDs can be supported without detection error. 
This coincides with Wainwright's analysis on the relationships among the number $M$ of observations, the problem dimension $N$, and the number $K_\mathrm{max}$ of nonzero elements of sparsity pattern reconstruction problem \cite{wainwright2009sharp}.
We can observe from (\ref{q_min}) that $\upsilon$ increases with the decrease of $N$ and the increase of $M$, $K$, and $\sigma^2$.
$\upsilon$ can be considered as a threshold to distinguish received pilot signals from noise and interference caused by the non-orthogonality of preamble sequences.
If a preamble arrives at the AP with received power lower than $\upsilon$, this preamble cannot be detected.

In the remaining part of this section, we first analyze the perfect AUD probability, i.e., the probability that Event 1 happens,
and then analyze the CE error when Events 1 or 2 happens,
based on the aforementioned LASSO problem model and the SG-based network model mentioned in Section II.

\subsection{Probability of Perfect Active User Detection }

As real-time IoT applications have high demands on low latency and high reliability,
the active MTCDs are expected to be detected within on shot transmission,
and thus Event 1 in Fig. \ref{venn_diagram} is expected to happen with high probability.
Therefore, we focus on the perfect AUD probability of CS-GF-NOMA in this subsection.

Based on Lemma 1, we can evaluate the probability of perfect AUD for CS-GF-NOMA as
\begin{align}\label{pr_succ_lasso1}
	\mathcal{P}_\text{PER} = \Pr \left\{ K \leq K_\text{max}, q_{\text{min}}>\upsilon \right\}
\end{align}

We introduce Theorem 1 to derive the closed-form expression of $\mathcal{P}_\text{PER}$, proved in Appendix B.

\begin{theorem}
The perfect AUD probability of the CS-GF-NOMA mMTC network with the SG network model is
\begin{align}\label{pr_succ_lasso_fin}
 \mathcal{P}_\mathrm{PER} =  \sum_{k=0}^{K_\mathrm{max}} \frac{e^{-\lambda}\lambda^k}{k!} \mathcal{P}_0^k,
\end{align}
where
\begin{align}\label{pr_succ_lasso_fin1}
\mathcal{P}_0 = 
	\frac{2 \left( {P}/{\upsilon^2}\right)^{\frac{2}{\alpha}} } {\alpha (D_1^2 - D_0^2)} 
	\left[ \Gamma\left(\frac{2}{\alpha},\frac{\upsilon^2 D_0^\alpha}{P} \right)
			-\Gamma\left(\frac{2}{\alpha},\frac{\upsilon^2D_1^\alpha}{P}\right) \right],
\end{align}
and $\Gamma(a,x) = \int_{x}^{\infty} t^{a-1}e^{-t} \mathrm{d}t$ is the upper incomplete Gamma function.

\end{theorem}

It is worth noting that $D_0$, $\alpha$, $\mathcal{P}_\mathrm{ACT}$, and $\sigma^2$ are usually not configurable in practical mMTC network.
From (\ref{pr_succ_lasso_fin}), the effect of the other configurable design parameters on $\mathcal{P}_\mathrm{PER}$ can be summarized as follows:
$\mathcal{P}_\mathrm{PER}$ increases as $M$ or $P$ increases and decreases
as $N$ or $D_1$ increases,
which will be verified by numerical results in Section \ref{Sec_Numerical}.

It is also worth noting that the ``perfect AUD" defined in $\mathcal{P}_\mathrm{PER}$, shown as Event 1 in Fig. \ref{venn_diagram}, is a strict AUD success from the view of the entire network,
which excludes both missed detection of active MTCDs and false detection of inactive MTCDs.
The successful detection probability of an active MTCD is positive correlated to but not equal to $\mathcal{P}_\mathrm{PER}$,
which is usually higher than $\mathcal{P}_\mathrm{PER}$ because some active MTCDs may still have chances to be detected when Events 2 or 3 happens.

\subsection{Error of Channel Estimation}

In this subsection, we analyze the CE error when Event 1 or 2 in Fig. \ref{venn_diagram} happens.
It is worth noting that accurate evaluation of the error of CS-based CE is difficult.
A commonly used tractable method is to approach the lower bound of the error by analyzing the error of the ideal oracle estimator \cite{davenport2012pros}, which has the perfect knowledge of signal sparsity profile.
Therefore, we first analyze the error of the oracle CE with the perfect knowledge of the successfully detected MTCDs.
Then by treating the number of detected MTCDs as a random variable and averaging the error of the oracle CE through this random variable, we obtain the closed-form expression of the average CE error in CS-GF-NOMA mMTC systems with the SG network model.

When both conditions (\ref{K_max_succ}) and (\ref{q_min_succ}) are satisfied, all the active MTCDs can be detected.
In this case, the CE error comes from the non-orthogonality of preambles and the noise.
When condition (\ref{K_max_succ}) is satisfied and condition (\ref{q_min_succ}) is not satisfied, the active MTCDs with AP received power lower than $\upsilon^2$ cannot be detected.
In this case, the CE error also comes from the interference of missed MTCDs,
besides the non-orthogonality of preambles and the noise.  
Therefore, we evaluate the CE error of the detected MTCDs by treating the preamble signals from the missed MTCDs as interference.

Let $\mathcal{S}_0 = \{ n\in\mathsf{Supp}(\mathbf{q}) \big| |q_n|\geq \upsilon \}$ and $\mathcal{S}_1 = \mathsf{Supp}(\mathbf{q}) \setminus \mathcal{S}_0$ denote the sets of successfully detected MTCDs and missed MTCDs, respectively.
The number of successfully detected MTCDs is $J = |\mathcal{S}_0|$.
The LS estimation of $\mathbf{q}_{\mathcal{S}_0}$ is
\begin{align}\label{ls_estimation}
	\widehat{\mathbf{q}}_{\mathcal{S}_0} & = \left( \mathbf{\Phi}^H_{\mathcal{S}_0} \mathbf{\Phi}_{\mathcal{S}_0} \right)^{-1}
	    \mathbf{\Phi}^H_{\mathcal{S}_0} \mathbf{y}_0 \notag\\
	  & = \mathbf{q}_{\mathcal{S}_0} + \left( \mathbf{\Phi}^H_{\mathcal{S}_0} \mathbf{\Phi}_{\mathcal{S}_0} \right)^{-1}
	    \mathbf{\Phi}^H_{\mathcal{S}_0} \left( \mathbf{\Phi}_{\mathcal{S}_1}\mathbf{q}_{\mathcal{S}_1} + \mathbf{w} \right).
\end{align}
Since the massive connectivity of mMTC is provided by preamble overload,
$\mathbf{\Phi}$ contains non-orthogonal columns,
and the component $\mathbf{\Phi}^H_{\mathcal{S}_0} \mathbf{\Phi}_{\mathcal{S}_1}$ in (\ref{ls_estimation}) cannot vanish.
We treat this unremovable component as interference to the estimation of $\mathbf{\Phi}_{\mathcal{S}_0}$.
Then we can evaluate the mean squared error (MSE) of the estimation as
\begin{align}
	\mathsf{MSE}_J & = \frac1{J} \mathbb{E}\left[
		\| \mathbf{q}_{\mathcal{S}_0} - \widehat{\mathbf{q}}_{\mathcal{S}_0} \|^2 \right] \notag\\
    & = \frac1{J} \mathbb{E}\left[ \left\|
    \left( \mathbf{\Phi}^H_{\mathcal{S}_0} \mathbf{\Phi}_{\mathcal{S}_0} \right)^{-1}
	    \mathbf{\Phi}^H_{\mathcal{S}_0} \left( \mathbf{\Phi}_{\mathcal{S}_1}\mathbf{q}_{\mathcal{S}_1} + \mathbf{w} \right)
    \right\|^2 \right].
\end{align}

As the complex Gaussian random matrix is adopted as the preamble matrix $\mathbf{\Phi}$,
a preliminary characterization of $\mathsf{MSE}_J$ can be obtained by the average over all possible preamble matrix realizations.
Further taking the the SG-model into consideration,
the closed-form expression of $\mathsf{MSE}_J$ is derived in Lemma 2, proved in Appendix C.

\begin{lemma}
    In the CS-GF-NOMA mMTC system with the SG network model,
    when $J$ active MTCDs ($J\leq M-4$) are successfully detected, the CE on the $J$ deteced MTCDs yields  MSE shown as Equation (\ref{eq_mse_thr}) at the top of the next page.
    \begin{figure*}[t!]
    	\begin{align}\label{eq_mse_thr}
    	\mathsf{MSE}_J = &\frac{\sigma^2}{M-J-1} + \frac{2 \lambda P \left(D_0^{2-\alpha}-D_1^{2-\alpha} \right)} 
    		{(\alpha-2) (M-J-1)  \left( D_1^2 - D_0^2 \right) } - \frac{2\lambda P^{\frac{2}{\alpha}}}
    			{\alpha (M-J-1) \left( D_1^2 - D_0^2 \right) {\upsilon}^{\frac{4}{\alpha}-2}}	\notag \\
    		& 
	    	\times \Bigg[ \Gamma\left( \frac{2}{\alpha} -1, \frac{\upsilon^2 D_0^\alpha}{P} \right) 
		    	- \Gamma\left( \frac{2}{\alpha} -1, \frac{\upsilon^2 D_1^\alpha}{P} \right)
		    	+ \Gamma\left( \frac{2}{\alpha}, \frac{\upsilon^2 D_0^\alpha}{P} \right) 
		    	- \Gamma\left( \frac{2}{\alpha}, \frac{\upsilon^2 D_1^\alpha}{P} \right) \Bigg].
    	\end{align}
    	\hrule
    \end{figure*}
\end{lemma}

Obviously, $J$ varies among different grant-free time slots and cannot be a priori acquired by the AP.
To achieve a generalized analysis on the SG network model and to provide some useful guidance on system design,
we evaluate the average CE error by averaging $\mathsf{MSE}_J$ through the distribution of $J$,
based on the SG network model.

With different setups of network deployment parameters,
i.e. with different values of $D_0$ or $\alpha$,
the AP received power and CE error may vary by several orders of magnitude.
To properly evaluate the accuracy of CE,
we evaluate the normalized mean squared error (NMSE) by normalizing the error with the actual value, which is defined as
\begin{align}
\mathsf{NMSE} & = \mathbb{E}\left[ \frac{\|\mathbf{q}_{\mathcal{S}_0}-\widehat{\mathbf{q}}_{\mathcal{S}_0}\|^2}
				{\|\mathbf{q}_{\mathcal{S}_0}\|^2 J } \right].
\end{align}

The average NMSE of CE in the CS-GF-NOMA mMTC system is given in Theorem 2, proved in Appendix D.

\begin{theorem}
	The average NMSE of CE in the CS-GF-NOMA mMTC system with the SG network model  is
	\begin{align}\label{th_2_nmse}
	\overline{\mathsf{NMSE}} = \frac1{\Xi}\sum_{j=1}^{K_\mathrm{max}} \mathsf{MSE}_j
	\sum_{k=j}^{K_\mathrm{max}} \frac{e^{-\lambda}\lambda^k}{k!} \tbinom{k}{j} \mathcal{P}_0^j (1-\mathcal{P}_0)^{k-j},
	\end{align}
	where the expectation $\Xi = \mathbb{E}\left[|{q}_n|^2\right]$ of the AP received power of a detected MTCD $n\in{\mathcal{S}_0}$ is given as (\ref{E_q0}).
\end{theorem}

\section{Performance Analysis for Data Transmission}

In this section, we evaluate the data transmission performance of the CS-GF-NOMA mMTC system in the data transmission phase.
We use achievable data rate \cite{shahini2019noma} as the metric to evaluate the data transmission performance, which is an important metric when concerning resource allocation and load balancing for multi-cell cases.
Higher achievable data rates indicate higher spectrum usage efficiency.

We consider the aggregate data rate when Event 1 or 2 happens, as shown in Fig. \ref{venn_diagram}.
In Event 1, every active MTCD can be detected.
Since the data symbols of different MTCDs are spread with unique preambles and superimposed on the whole grant-free sub-channels with different AP received powers,
successive interference cancelation (SIC) can be adopted for data decoding.
With SIC, the data decoding of the MTCDs with higher AP received power will always precede that of the MTCDs with lower received power,
and the data signals of the MTCDs with lower received powers are treated as interference during the data decoding of the MTCDs with higher received powers.
In Event 2, a part of the active MTCDs are detected, and the other active MTCDs are missed detection.
The data signals of the missed detected MTCDs are treated as interference through the whole SIC decoding procedure of the successfully detected MTCDs.
We do not consider Event 3 for the following two reasons:
first, the recovered sparsity profile in this case is a mixture of missed detection and false detection, the performance of which is difficult to analyze;
second, in practical system design, the probability of Event 1 is expected to be high enough that the probability of Event 3 is very small, and thus the performance in this case have small influence to the overall network performance.

First, we derive the achievable aggregate rate given that $J$ MTCDs are successfully detected among $K$ active MTCDs. 
Suppose that the AP received powers of the $K$ active MTCDs are ranked such that
$\left| q_{n_1}\right| \geq \cdots \geq \left| q_{n_J}\right| \geq 
 \upsilon >  \left| q_{n_{J+1}}\right| \geq \cdots \geq \left| q_{n_K}\right| $,
 where $n_k$ denotes the index of the $k$th active MTCD in $\mathcal{N}$.
During each data symbol $l$, the AP decodes the data symbols of $J$ detected MTCDs successively from $s_{n_1,l}$ to $s_{n_J,l}$ with SIC.
When decoding $s_{n_j,l}$, the received signals of $s_{n_{k},l}$ with $J<k\leq K$ are treated as interference.
Therefore, the achievable rate of the $j$th detected MTCD is \cite{shahini2019noma}
\begin{align}
	R_j = \log_2\left( 1+\frac{\left| q_{n_j} \right|^2}
        {\sum_{k=j+1}^{K} \left| q_{n_k} \right|^2 + M \sigma^2} \right),
\end{align}
and the aggregate data rate of the $J$ detected MTCDs is
\begin{align}\label{R_KJ}
	{R}_{K,J} = \sum_{j=1}^{J} R_j
      = \log_2 \left( \frac{\sum_{k=1}^{K} \left| q_{n_k} \right|^2 + M \sigma^2}
          {\sum_{j=J+1}^{K} \left| q_{n_j} \right|^2 + M \sigma^2} \right).
\end{align}

Next, we generalize the aggregate rate (\ref{R_KJ}) to the SG-based network model by averaging ${R}_{K,J}$ through the distribution of $J$, $K$, channel gains, and noise powers.
Theorem 3 gives the expression of the average aggregate data rate of GF-NOMA mMTC systems, proved in Appendix E.

\begin{theorem} \label{Th_average_rate}
	The average aggregate data rate $\overline{R}_\mathrm{A}$ of the CS-GF-NOMA mMTC system with the SG network model is given as Equation (\ref{ave_R}) at the top of the next page.
	\begin{figure*}[!t]
		\begin{align}\label{ave_R}
		\overline{R}_\mathrm{A} \geq & \frac1{\ln 2} \sum_{k=1}^{K_\mathrm{max}} \frac{e^{-\lambda}\lambda^k}{k!} \int_{0}^{\infty}
			\frac{e^{-s}}{s}\left[ 1- Q^k(s) \right] \mathrm{d}s +\log_2\left(M \sigma^2\right) - 
			\log_2 \Bigg(  \frac{2 \lambda P}{D_1^2 - D_0^2} \Bigg\lbrace \frac{D_0^{2-\alpha} - D_1^{2-\alpha}}{\alpha-2}
			 \notag \\
			& - \frac1{\alpha} \left( \frac{P}{\upsilon^2}\right)^{\frac{2}{\alpha}-1} \Bigg[
				\Gamma\left( \frac{2}{\alpha} -1, \frac{\upsilon^2 D_0^\alpha}{P} \right) 
				- \Gamma\left( \frac{2}{\alpha} -1, \frac{\upsilon^2 D_1^\alpha}{P} \right)
				+ \Gamma\left( \frac{2}{\alpha}, \frac{\upsilon^2 D_0^\alpha}{P} \right) 
				- \Gamma\left( \frac{2}{\alpha}, \frac{\upsilon^2 D_1^\alpha}{P} \right) \Bigg] \Bigg\rbrace
			+M \sigma^2 \Bigg),
		\end{align}
		where $Q(s) = \frac{2M \sigma^2}{sP(D_1^2-D_0^2)(\alpha+2)}
				\left[ D_1^{\alpha+2} G_\alpha \left( -\frac{M \sigma^2 }{sP} D_1^\alpha\right)
					-D_0^{\alpha+2} G_\alpha \left( -\frac{M \sigma^2}{sP} D_0^\alpha \right)\right]$,
		and $G_\alpha (x) = F\left(1, \frac{2}{\alpha}+1;\frac{2}{\alpha}+2;x \right)$ is the Gauss hypergeometric function \citep[Section 9.111]{izrail2007table}.
		\vspace{3pt}
		\hrule
	\end{figure*}
\end{theorem}

Equation (\ref{ave_R}) can be evaluated via well-known mathematical packages, as Matlab or Mathematica \cite{yang2017uplink}.
$Q(s)$ monotonically increases with $s$ and $\lim_{s \to \infty}Q(s)=0$.
Therefore, the integral in (\ref{ave_R}) can be separated into two segments such that 
$\int_{0}^{\infty} \frac{e^{-s}}{s}\left[ 1- Q^k(s) \right] \mathrm{d}s = \int_{0}^{s_0} \frac{e^{-s}}{s}\left[ 1- Q^k(s) \right] \mathrm{d}s + \int_{s_0}^{\infty} \frac{e^{-s}}{s}\left[ 1- Q^k(s) \right] \mathrm{d}s $,
where $Q(s_0) = \epsilon$ for a small enough $\epsilon>0$.
The first segment can be evaluated via numerical integration methods.
The second segment can be approximated with $\mathsf{Ei}(-s_0)$, 
where $\mathsf{Ei}(x)=\int_{-\infty}^{x} \frac{e^t}{t} \mathrm{d}t$ is the exponential integral function.

\begin{figure*}
	\centering
	\subfigure[]{\includegraphics[scale=0.64]{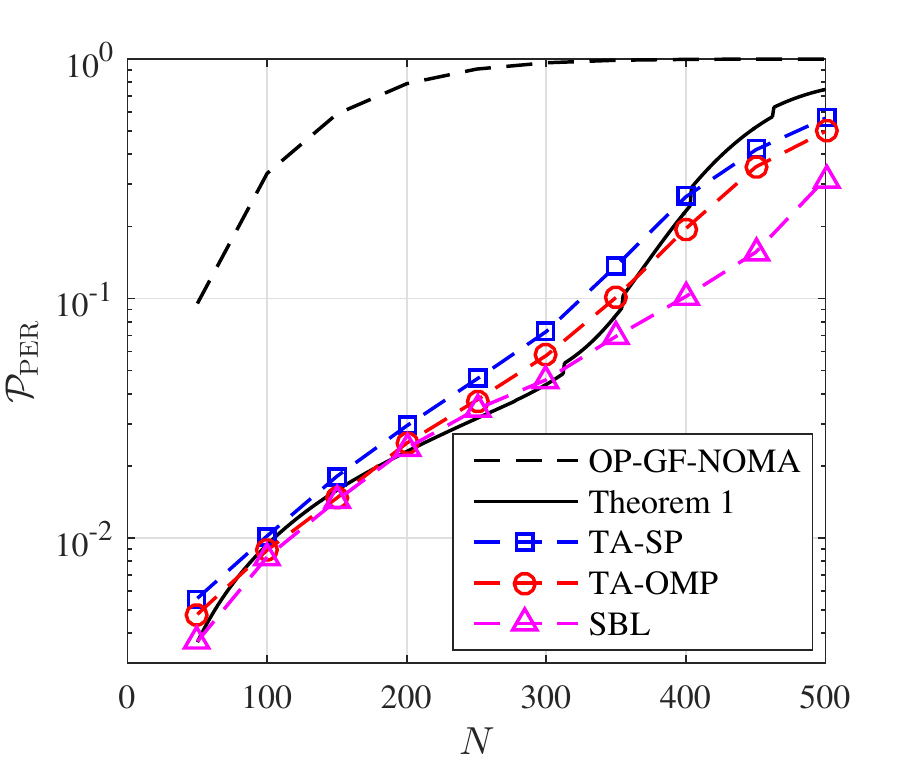}}
	\subfigure[]{\includegraphics[scale=0.64]{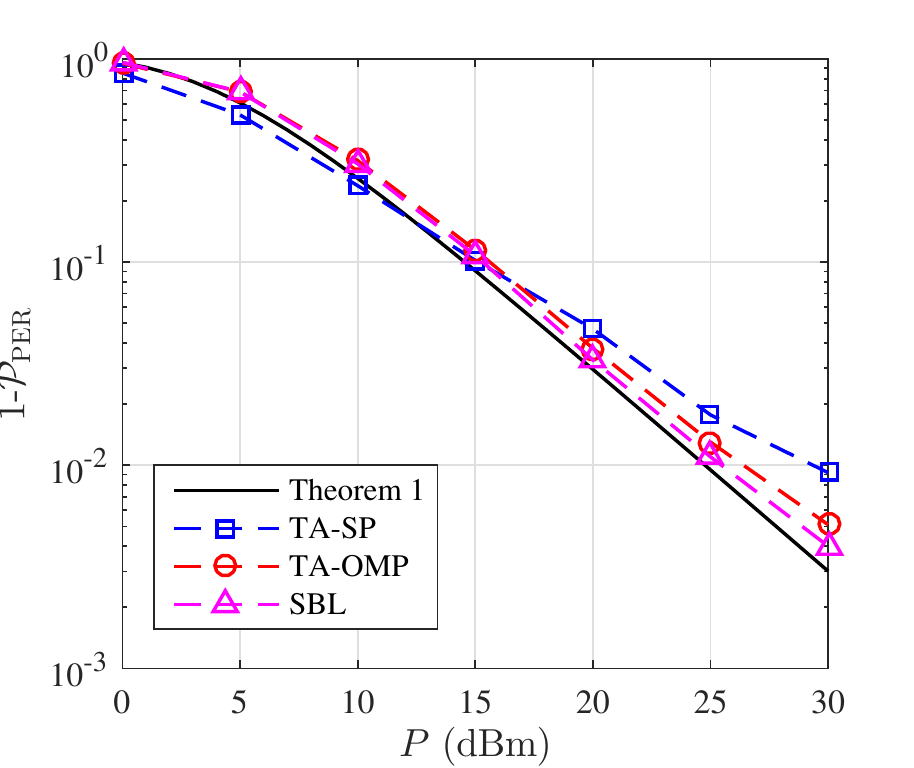}}
	\subfigure[]{\includegraphics[scale=0.64]{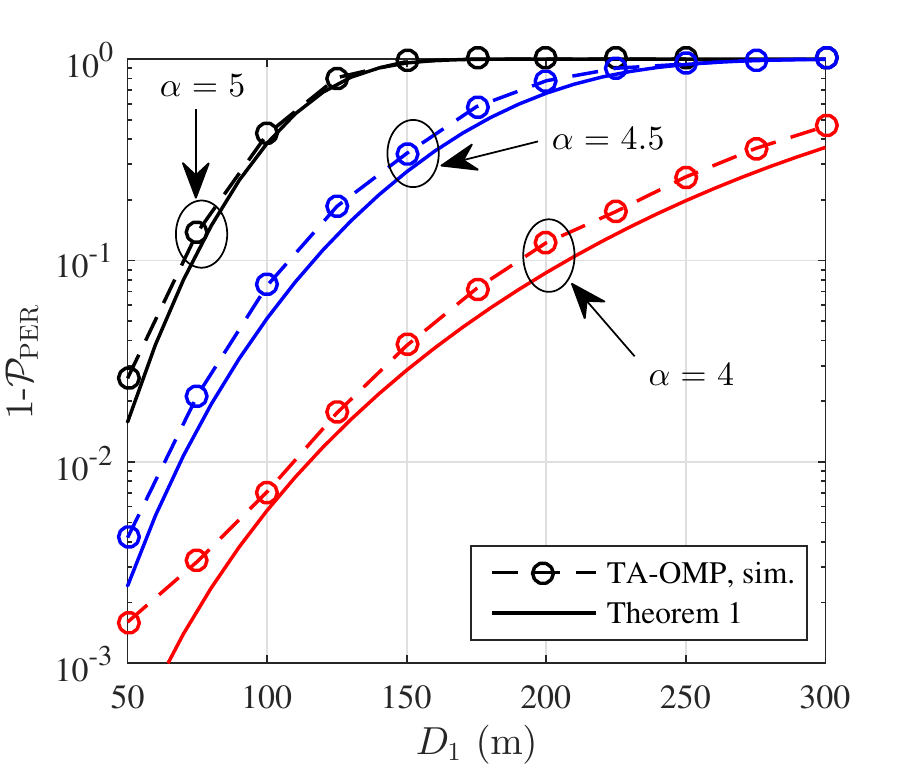}}
	\caption{Perfect AUD success probability of the CS-GF-NOMA mMTC system. Analytical result is compared with the simulation results of CB-OP, TA-SP, TA-OMP, and SBL algorithms. (a) $1-\mathcal{P}_\mathrm{PER}$ versus $N$. (b) $1-\mathcal{P}_\mathrm{PER}$ versus $P$. (c) $1-\mathcal{P}_\mathrm{PER}$ versus $D_1$. }
	\label{fig_failure_N}
\end{figure*}

\section{Numerical Results and Discussion}\label{Sec_Numerical}

In this section, we evaluate the performance of CS-GF-NOMA via Monte Carlo simulations
and verify the accuracy of the mathematical performance analysis obtained in the previous sections with the simulation results. 
The complex Gaussian random matrix is used as the preamble matrix $\mathbf{\Phi}$.
Table \ref{Sim_Para} shows the typical values of simulation parameters.

\begin{table}[h]
	\caption{Simulation parameters}
	\label{Sim_Para}\centering
	\small
	\begin{tabular}{ll} 
		\hline
		\textbf{Parameter} & \textbf{Value} \\ \hline
		Cell radius $D_0$, $D_1$ & 10 m, 150 m \\
		Number of potential MTCDs $N$ & 240 \\
		Preamble length $M$ & 120 \\
		Noise power on each subchannel $\sigma^2$ & $-110$ dBm\\
		Pathloss exponent $\alpha$ & 4 \\
		MTCD active probability $\mathcal{P}_\mathrm{ACT}$ & 0.1 \\
		MTCD static circuit power $P_\mathrm{S}$ & 3 mW\\
		MTCD dynamic circuit power $P_\mathrm{D}$ & 100 mW\\
		MTCD transmit power $P$ & 20 dBm \\ 
		MTCD antenna efficiency $\varepsilon$ & 0.5 \\
		\hline
	\end{tabular}
\end{table}

\subsection{Performance Evaluation of CS-based AUD} 

Fig. \ref{fig_failure_N} verifies our theoretical analysis on the performance of CS-GF-NOMA networks via Monte Carlo simulations and makes a comparison between CS-GF-NOMA and OP-GF-NOMA.
Figs. \ref{fig_failure_N}(a) and (b) illustrate $\mathcal{P}_\mathrm{PER}$ of the GF-NOMA network versus $N$ and $P$, respectively. 
Fig. \ref{fig_failure_N}(c) illustrates $\mathcal{P}_\mathrm{PER}$ of the GF-NOMA network versus $D_1$ for different $\alpha$.
The values of the other system parameters refer to Table \ref{Sim_Para}.
To show the upper area of $\mathcal{P}_\mathrm{PER}$ more clearly, we show $1-\mathcal{P}_\mathrm{PER}$ with logarithmic axis.

In OP-GF-NOMA, Zadoff-Chu sequences are used as preambles.
For fair comparison, $\mathcal{P}_\mathrm{PER}$ denotes the probability that all the active MTCDs can be successfully detected without any preamble collision, false detection, or missed detection.
From Fig. \ref{fig_failure_N}(a), CS-GF-NOMA has better AUD performance than OP-GF-NOMA.
This is because preamble collision happens very frequently in OP-GF-NOMA,
while CS-GF-NOMA can effectively avoid the collision with overloaded preambles being pre-assigned to the MTCDs.

We verify the analytical results from Theorem 1 with the simulation results of some state-of-the-art CS-based MUD algorithms for GF-NOMA,
which include the threshold-based subspace pursuit algorithm (TA-SP) \cite{du2018block}, the threshold-based orthogonal matching pursuit algorithm (TA-OMP) \cite{liu2017blind}, and the sparse Bayesian learning algorithm (SBL) \cite{zhang2018block}.
It is worth noting that the original LASSO problem (\ref{lasso}) is NP-hard, direct solution of which is with prohibitive complexity.
Orthogonal matching pursuit (OMP) and subspace pursuit algorithm (SP) are two classic greedy algorithms that achieve the sub-optimal solution to the LASSO problem with much lower complexity.
However, OMP and SP require the prior information on the sparsity, i.e. the number of active MTCDs, which is unpractical in the CS-GF-NOMA mMTC system.
TA-OMP and TA-SP are two improved algorithms to settle the issue of unknown number of active MTCDs by using sparsity-related thresholds to control the iteration procedure.
From Fig. \ref{fig_failure_N}, Theorem 1 provides good approximation of the actual AUD performance of  TA-SP and TA-OMP.
The simulation results of SBL is obviously better than the analytical results when perfect AUD probability is lower than $0.9$.
However, our analysis still offers good guidance for system design because the reliability-sensitive IoT applications generally require a $\mathcal{P}_\mathrm{PER}$ at least $0.9$.
Theorem 1 is better at represent the performance of the greedy TA-SP and TA-OMP alogrithms because the $\ell_1$-panalty LASSO problem model (\ref{lasso}) has a greedy nature, which tries to recover $\widehat{\mathbf{q}}$ with as few non-zero entries as possible and thus gives priority to the large entries in $\mathbf{q}$.

It is worth noting that Fig. \ref{fig_failure_N}(a) shows the inevitable quantization error of the analytical results: there are some discontinuity points where $1-\mathcal{P}_\mathrm{PER}$ suddenly increase with the increase of $N$.
This is because the upper limit  $K_\mathrm{max}$ of the summation in (\ref{pr_succ_lasso_fin}) is an integer, which does not change continuously with $N$ according to (\ref{K_max}).

In practical network design, $\mathcal{P}_\mathrm{PER}$ should be at least $0.9$ to guarantee AUD reliability.
In this regard, $N$ should be at most 355, 350, 325, and 400 for Theorem 1, TA-OMP, TA-SP, and SBL algorithms, respectively from Fig. \ref{fig_failure_N}(a).
$P$ should be at least 14.5 dBm, 15.6 dBm, 15.2 dBm, and 15.4 dBm for Theorem 1, TA-OMP, TA-SP, and SBL algorithms, respectively from Fig. \ref{fig_failure_N}(b).

From Fig. \ref{fig_failure_N}(c), the AUD performance of GF-NOMA can be improved with the increase of $D_1$ and $\alpha$. 
The reason is obvious: longer link distances and higher $\alpha$ indicate worse channel conditions.
Therefore, under the condition of constant total number $N$ of potential MTCDs in the cell, $D_1$ is expected to be as small as possible to achieve better AUD performance.
We will consider another scenario in the next subsection under the condition of constant device deployment density and find the optimal $D_1$ maximizing the APCE.
Moreover, Theorem 1 can provide better approximation of the Monte Carlo simulation results with larger $\alpha$.

\begin{figure}[t]
	\centering
	\includegraphics[scale=0.61]{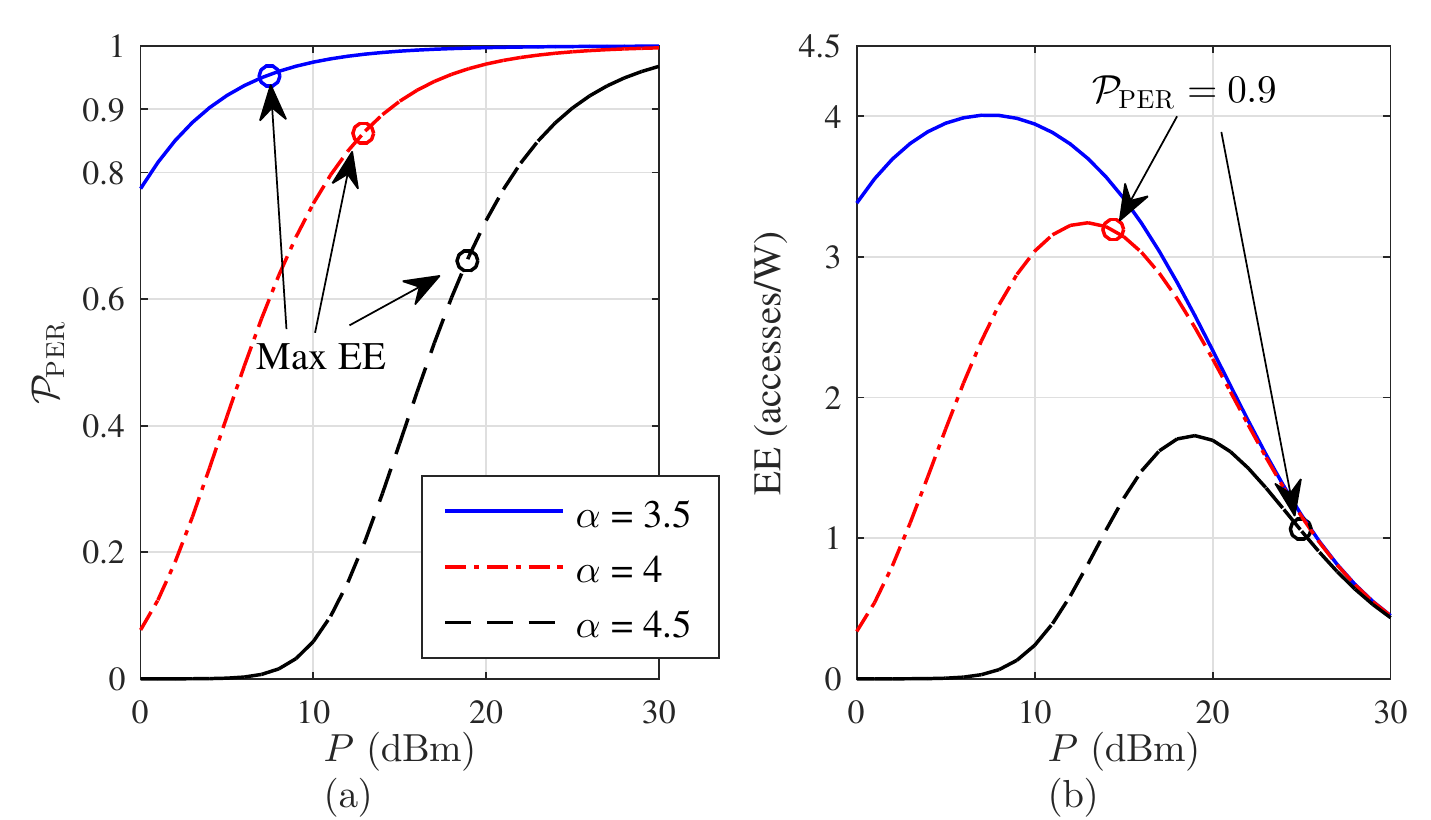}
	\caption{EE optimization of the CS-GF-NOMA mMTC system. (a) $\mathcal{P}_\mathrm{PER}$ versus $P$ for different values of $\alpha$. (b) EE versus $P$ for different values of $\alpha$.}
	\label{fig_ee_ptx_double}
\end{figure}

\subsection{Optimization of EE and APCE}

In this subsection, we consider two performance metrics related to $\mathcal{P}_\mathrm{PER}$,
i.e. the EE and the APCE,
which are important metrics for system design and network deployment.
For each metric, we first consider unconstrained optimization and then consider AUD reliability-constrained optimization.

\subsubsection{Energy Efficiency}

By analyzing Fig. \ref{fig_ee_ptx_double}, we optimize the EE of the CS-GF-NOMA mMTC system.
The EE is defined as the ratio of the average number of stably detected MTCDs versus the average total power consumption of all the MTCDs in the system, i.e. $\mathsf{EE} = \frac{\lambda \mathcal{P}_\text{PER}}{ N \overline{P}_\mathrm{DEV}}$.
Fig. \ref{fig_ee_ptx_double}(a) and (b) respectively illustrate the analytical results of $\mathcal{P}_\mathrm{PER}$ and the EE of the GF-NOMA system versus $P$ for different values of $\alpha$. 
The values of the other system parameters refer to Table \ref{Sim_Para}.

From Fig. \ref{fig_ee_ptx_double}(a), $\mathcal{P}_\mathrm{PER}$ increases with the increase of $P$.
Meanwhile, the total power assumption of MTCDs also increases.
Therefore, the EE observed from Fig. \ref{fig_ee_ptx_double}(b) first increases and then decreases.
There is an optimal $P$ to balance the two effects and then to achieve a maximum EE.
One-dimensional optimization algorithms \cite{antoniou2007practical} can be adopted to obtain the optimal $P$.

Using golden-section search \citep[Section 4.4]{antoniou2007practical}, the optimal $P$ maximizing the EE are $7.4$ dBm, $12.9$ dBm, and $18.9$ dBm when $\alpha = 3.5$, $4$, and $4.5$, respectively.
The corresponding EE are $4.01$ accesses/W, $3.24$ accesses/W, and $1.73$ accesses/W,
and the corresponding $\mathcal{P}_\mathrm{PER}$ are $0.95$, $0.86$, and $0.66$, respectively,
marked with circle markers in Fig. \ref{fig_ee_ptx_double}(a).

Obviously, $\mathcal{P}_\mathrm{PER}$ yielded by EE optimization in the case of $\alpha = 4$ and $4.5$ is relatively low, with the simulation setups in this paper.
We choose the lowest $P$ that yields $\mathcal{P}_\mathrm{PER}$ at least $0.9$ to guarantee AUD reliability.
In this way, when $\alpha = 4$ and $4.5$, the AUD reliability-constrained optimal $P$ are $14.5$ dBm and $24.9$ dBm,
which yield the EE of $3.19$ accesses/W and $1.06$ accesses/W, respectively,
marked with circle markers in Fig. \ref{fig_ee_ptx_double}(b).

\subsubsection{Access Point Coverage Efficiency}

\begin{figure}[t]
	\centering
	\includegraphics[scale=0.61]{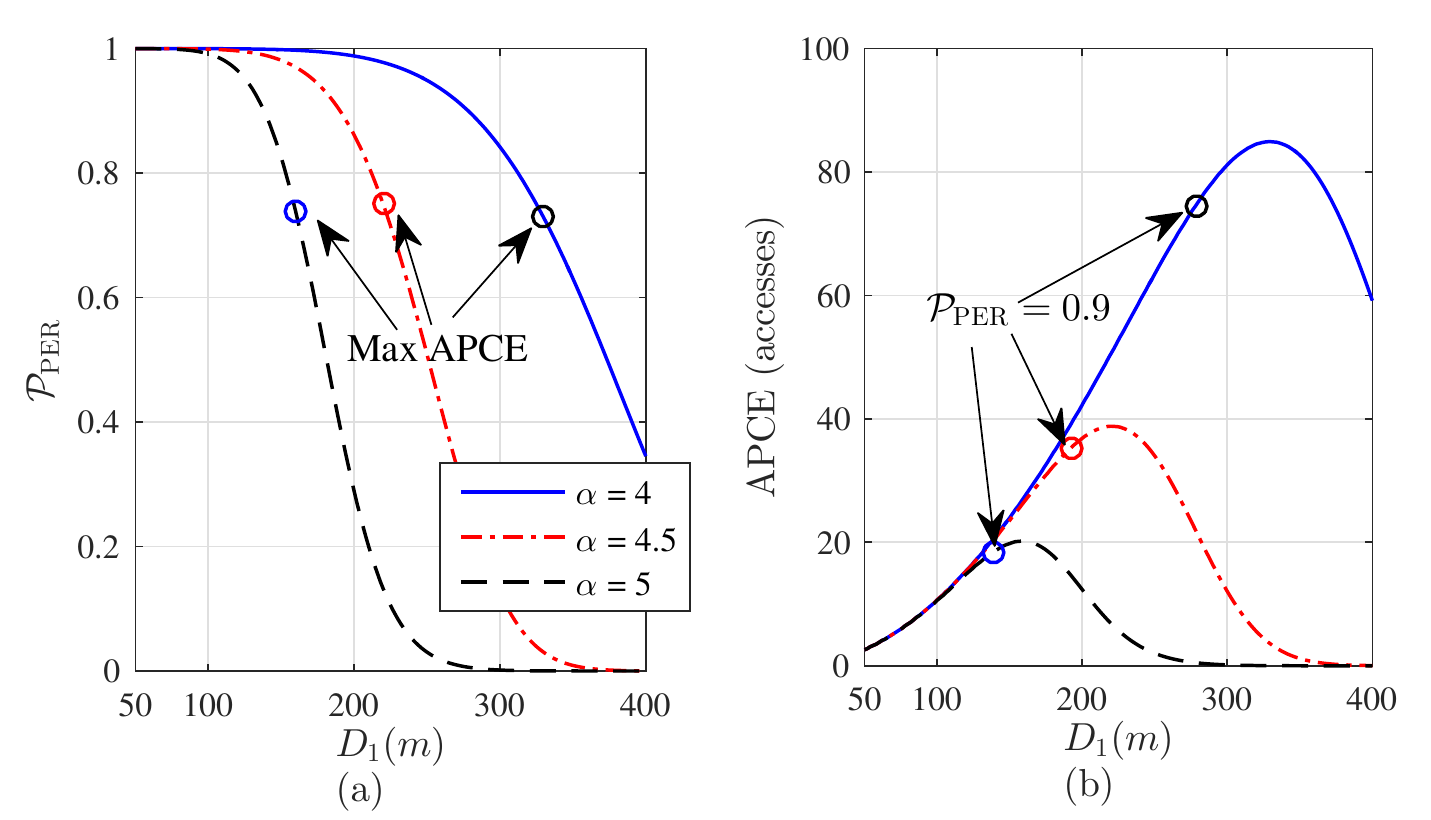}
	\caption{APCE Optimization of the GF-NOMA mMTC system with $N = \lambda_0 \pi (D_1^2 - D_0^2)$. (a) $\mathcal{P}_\mathrm{PER}$ versus $D_1$ for different values of $\alpha$. (b) APCE versus $D_1$ for different values of $\alpha$.}
	\label{fig_connection_D0_double}
\end{figure}

\begin{figure*}
	\centering
	\subfigure[]{\includegraphics[scale=0.62]{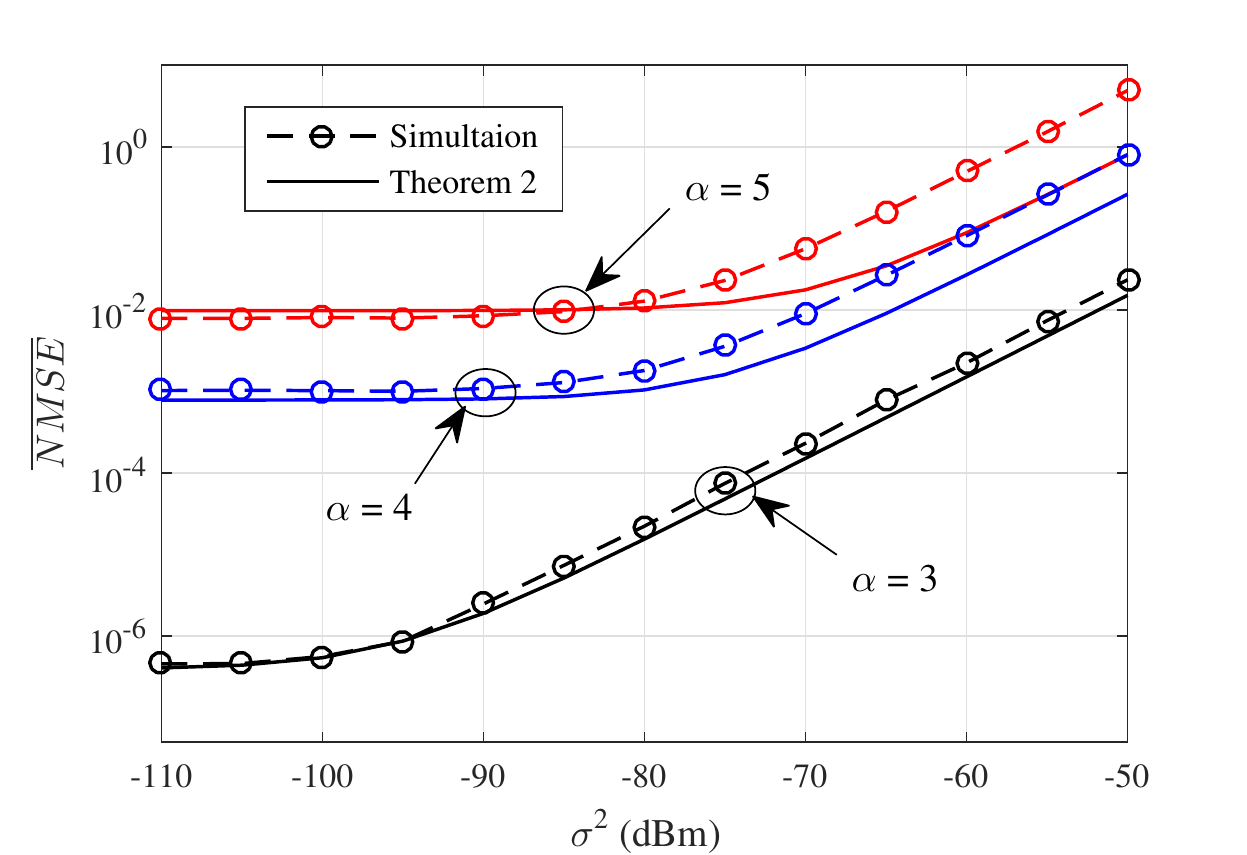}}\hspace{20pt}
	\subfigure[]{\includegraphics[scale=0.62]{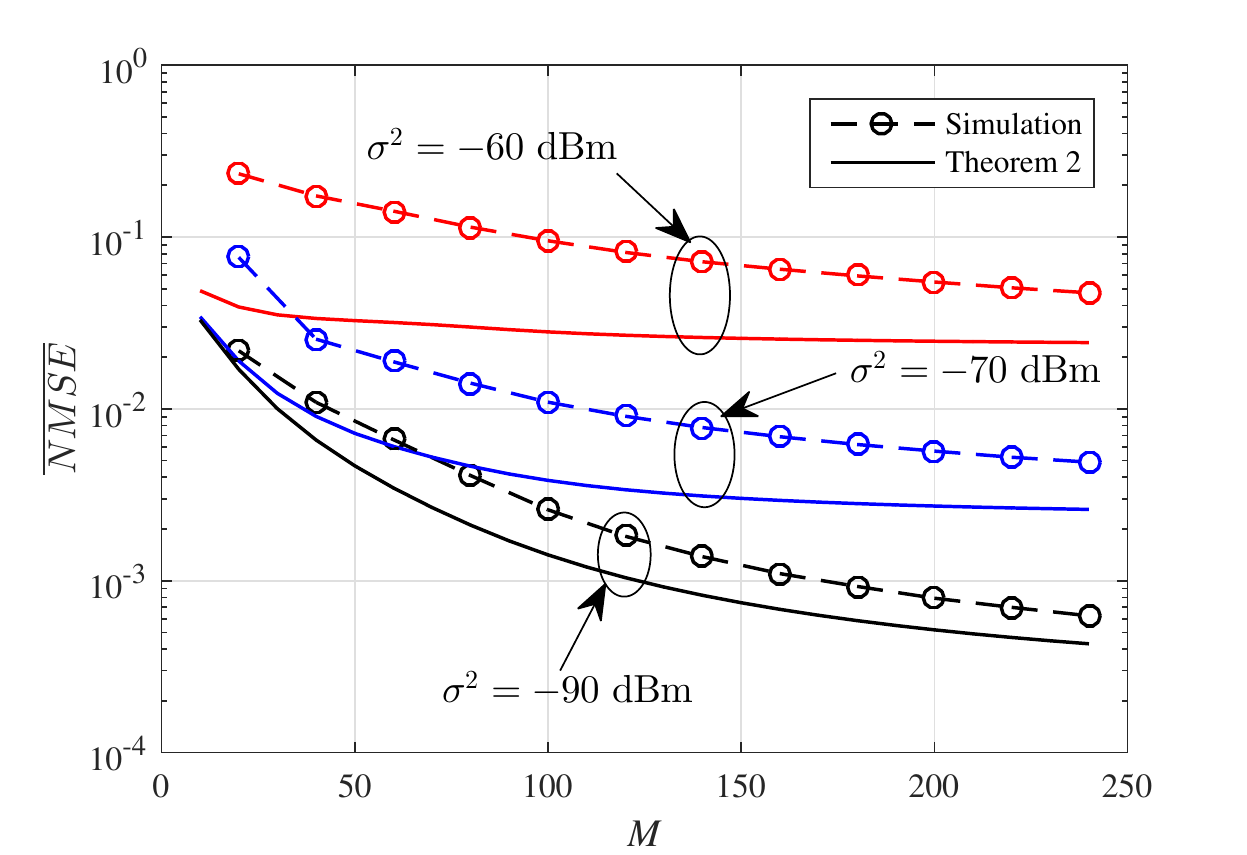}}
	\caption{ CE error of the CS-GF-NOMA mMTC system.
		(a) Average NMSE versus $\sigma^2$ for different values of $\alpha$. 
		(b) Average NMSE versus $M$ for different values of $\sigma^2$. }
	\label{fig_CE_error}
\end{figure*}

By analyzing Fig. \ref{fig_connection_D0_double}, we optimize the APCE of the CS-GF-NOMA mMTC system.
It is worth noting that Fig. \ref{fig_connection_D0_double} considers another scenario with constant device deployment density, 
where the total number of potential active MTCDs connected to an AP is proportional to the coverage area of this AP,
i.e.  $N = \lambda_0 \pi (R_1^2 - R_0^2)$, and $\lambda_0$ is the MTCD deployment density.
We set $\lambda_0 = 3410 ~ \mathrm{devices/km}^2$,
corresponding to the previous simulation setup that $N=240$ potential MTCDs are distributed in the cell of $D_0 = 10~\mathrm{m}$ and $D_1 = 150~\mathrm{m}$.
Average number of stable accesses supported by the AP can be evaluated as $\lambda_0 \pi (R_1^2 - R_0^2) \mathcal{P}_\mathrm{PER}$.
This optimization answers the following question on network deployment: 
in a large area with a large number of MTCDs, which requires several APs to cover the whole area,
how large the coverage area of each AP should be or how many APs should be deployed to achieve the most efficient usage of each AP?
Fig. \ref{fig_connection_D0_double}(a) and (b) respectively illustrate the analytical results of $\mathcal{P}_\mathrm{PER}$ and the APCE of CS-GF-NOMA systems versus $D_1$, with different values of $\alpha$. 
The values of the other system parameters refer to Table \ref{Sim_Para}.

From Fig. \ref{fig_connection_D0_double}(a), $\mathcal{P}_\mathrm{PER}$ decreases with the increase of $D_1$ because the average path loss of the MTCDs gets more serious.
Meanwhile, the average number of active MTCDs within each time frame increases.
Therefore, the APCE observed from Fig. \ref{fig_connection_D0_double}(b) first increases and then decreases.
There is an optimal cell size to balance the two effects and then to achieve a maximum APCE.
One-dimensional optimization algorithms \cite{antoniou2007practical} can be adopted to obtain the optimal $P$.

Using golden-section search \citep[Section 4.4]{antoniou2007practical}, the optimal $D_1$ maximizing the APCE are $328.4$ m, $221.7$ m, and $159.5$ m when $\alpha = 4$, $4.5$, and $5$, respectively.
The corresponding APCE are $84.9$, $38.5$, and $20.3$ accesses,
and the corresponding $\mathcal{P}_\mathrm{PER}$ are $0.73$, $0.73$, and $0.74$, respectively,
marked with circle markers in Fig. \ref{fig_connection_D0_double}(a).

Obviously, $\mathcal{P}_\mathrm{PER}$ yielded by APCE optimization is relatively low.
We choose the lowest $P$ that yields $\mathcal{P}_\mathrm{PER}$ at least $0.9$ to guarantee AUD reliability.
In this way, when $\alpha = 4$, $4.5$, and $5$, the AUD reliability-constrained optimal $D_1$ are $278.1$ m, $193.2$ m, and $138.3$ m,
which yield APCE of $75.2$ accesses, $35.5$ accesses, and $18.4$ accesses, respectively,
marked with circle markers in Fig. \ref{fig_connection_D0_double}(b).

\subsection{NMSE of Channel Estimation} 

Fig. \ref{fig_CE_error} illustrates the CE error of the CS-GF-NOMA mMTC system,
where Fig. \ref{fig_CE_error}(a) illustrates the NMSE versus $\sigma^2$ for different values of $\alpha$,
and Fig. \ref{fig_CE_error}(b) illustrates the NMSE versus $M$ for different values of $\sigma^2$.

From Fig. \ref{fig_CE_error}(a), Theorem 2 provides good approximation for CE error, and the gap between the analytical results and the simulation results shrinks with the decrease of $\sigma^2$.
Moreover, the NMSE decreases with the decrease of $\sigma^2$,
and there is a lower bound for the {NMSE}.
When $\sigma^2$ is small, the CE error is dominated by the interference from missed MTCDs and the non-orthogonality of preambles.
As the figure shows, the two effects are well evaluated by our analysis.

From Fig. \ref{fig_CE_error}(b), the gap between the analytical results of Theorem 2 and the simulation results shrinks with the increase of $M$.
Moreover, the NMSE decreases with the increase of $M$, especially when $M$ is small.
However, when $M$ is large, the channel estimation accuracy improved by increasing $M$ is not evident, especially when $\sigma^2$ is large.

\subsection{Average Aggregate Data Rate}

\begin{figure*}
	\centering
	\subfigure[]{\includegraphics[scale=0.59]{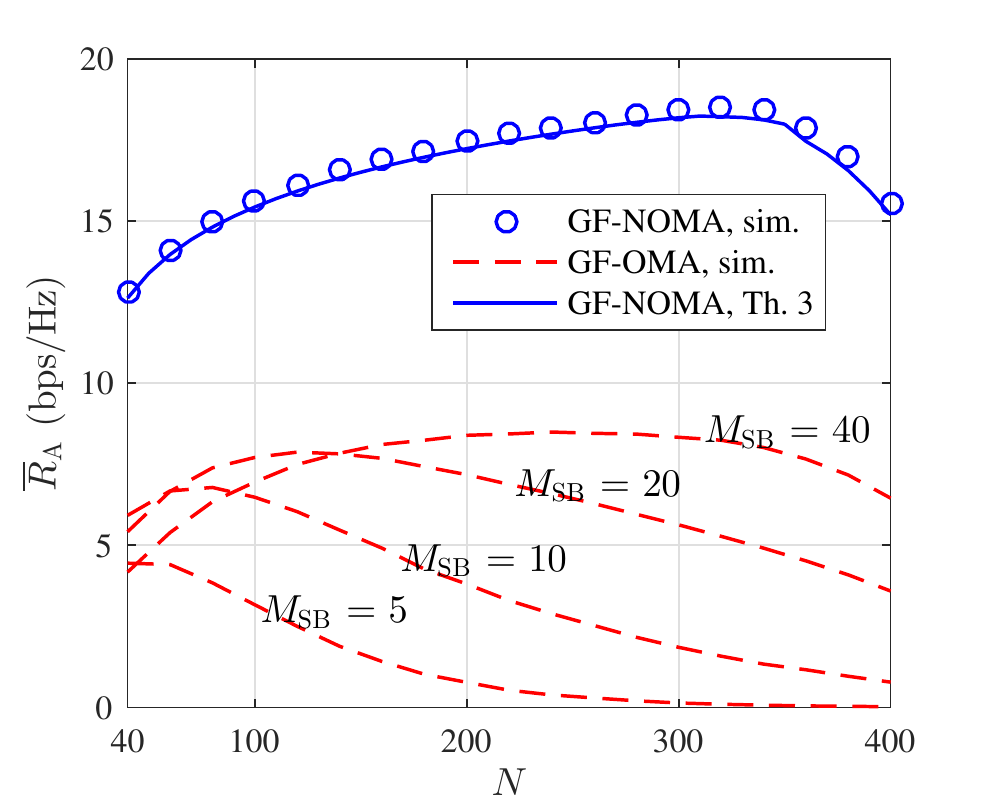}}
	\subfigure[]{\includegraphics[scale=0.59]{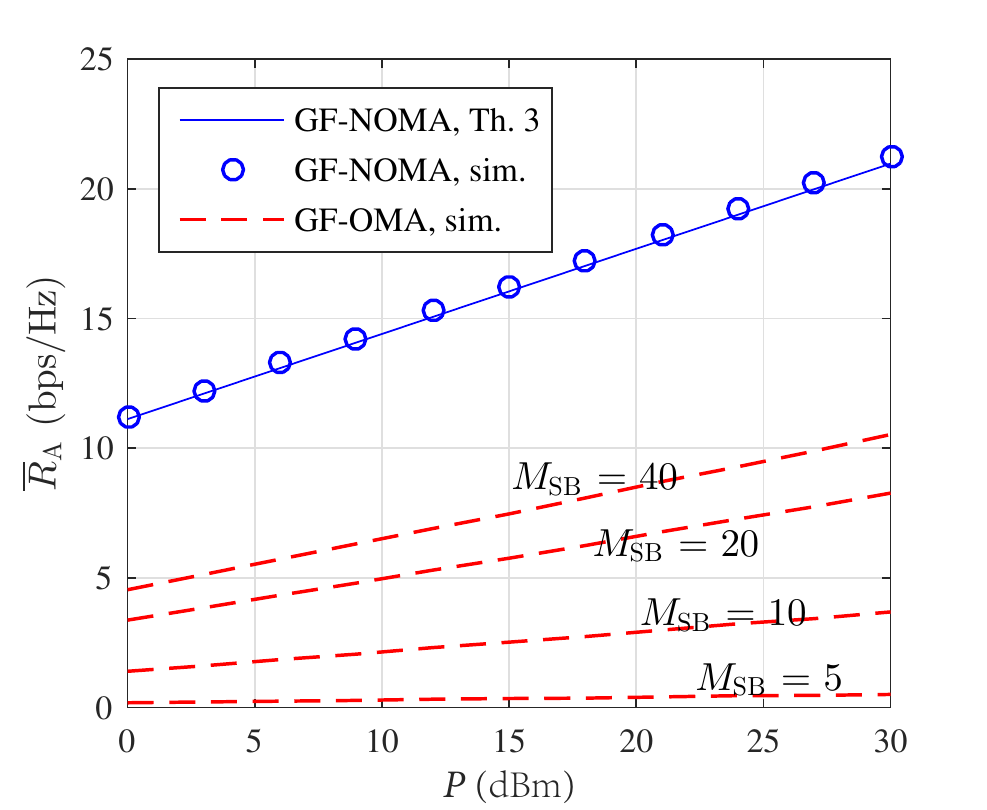}}
	\subfigure[]{\includegraphics[scale=0.59]{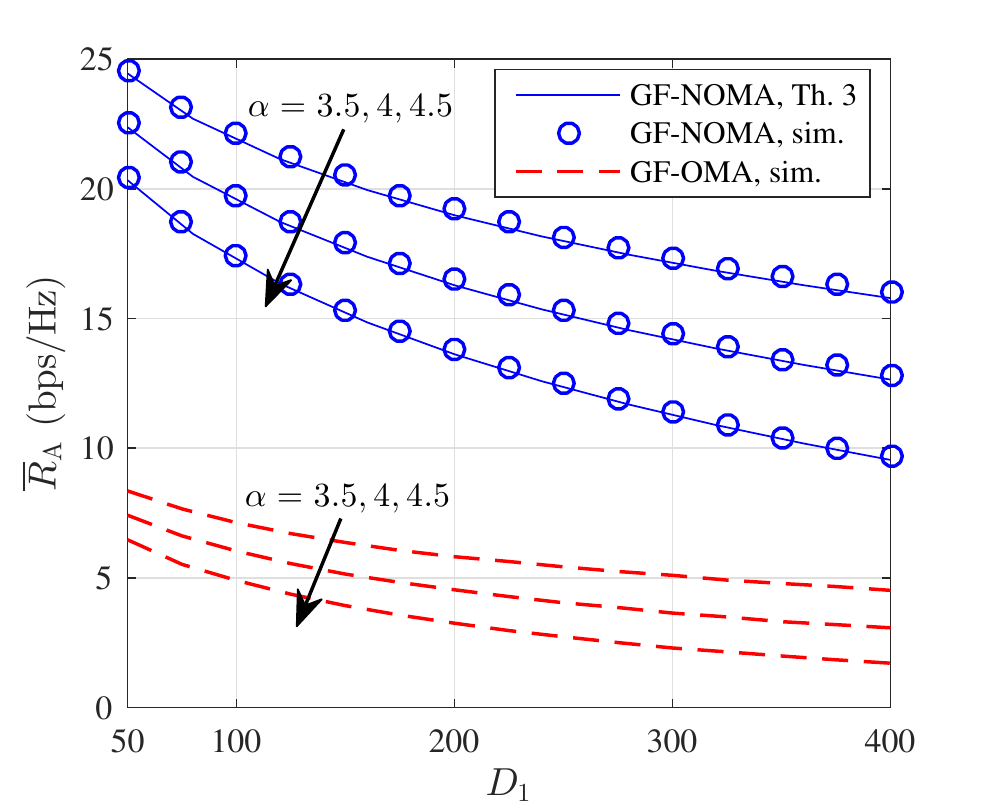}}
	\caption{Average aggregate data rate of the CS-GF-NOMA mMTC system, compared with the GF-OMA mMTC system. (a) $\overline{R}_\mathrm{A}$ versus $N$. (b) $\overline{R}_\mathrm{A}$ versus $P$. (c) $\overline{R}_\mathrm{A}$ versus $D_1$ for different values of $\alpha$. }
	\label{fig_Ra}
\end{figure*}

Fig. \ref{fig_Ra} illustrates the average aggregate data rate of the CS-GF-NOMA mMTC system.
We make a comparison between GF-NOMA and GF-OMA mMTC.
GF-OMA uses the same preamble sequences that GF-NOMA uses for MUD and CE
and divide the total $M$ sub-channels into $M_\mathrm{SB}$ sub-bands for data transmission.
Each MTCD is designated to a sub-band at the initial of the network.
When an MTCD is active, it will directly transmit its preamble over the $M$ sub-channels and its data over its designated sub-band.
It is worth noting that since multiple MTCDs are designated to one sub-band, there may be the case that more than one MTCDs transmit their data on the same sub-band, which indicates the happening of collision.
We assume that if collision happens among the successfully detected MTCDs by CS-based AUD, then the data of these MTCDs are not decodable.
On the other hand, if collision happens between one successfully detected MTCD and one or more missed MTCDs, then the data of the successfully detected MTCD is still decodable by treating the data signals from the missed MTCDs as interference.
This assumption is valid because the received power of a missed detected MTCD is naturally low.

It can be observed from the figures that GF-NOMA yields much higher average aggregate data rate than GF-OMA. 
The reason is two-fold:
first, the SIC receiver of GF-NOMA is able to decode the data signals in collision;
second, the superimposed structure of NOMA signals helps to improve the spectrum efficiency. 
From Fig. \ref{fig_Ra}(a), the average aggregate data rates of GF-NOMA and GF-OMA first increases and then decreases.
For GF-NOMA, the reason is that although more active MTCDs exist with larger $N$, the AUD failure probability also increases, which causes more missed detection.
For GF-OMA, the decrease of the average aggregate data rate comes from both the increase of the AUD failure probability and the increase of the collision probability.
From Figs. \ref{fig_Ra}(a) and (b), GF-OMA yields higher aggregate data rate with a higher $M_\mathrm{SB}$ because collision happens with lower probability with more sub-bands.

From Fig. \ref{fig_Ra}(c), the average aggregate data rates of GF-NOMA and GF-OMA decrease with the increase of $D_1$ or the increase of $\alpha$.
The reason is obvious: longer link distances and higher $\alpha$ indicate worse channel conditions.

\section{Conclusions}

In this paper, we considered the modeling, analysis, and optimization of the CS-GF-NOMA mMTC system for IoT applications.
We proposed an analytic model of the CS-GF-NOMA mMTC system to realize tractable analysis, where the SG model was adopted to formulate the network deployment, and the LASSO model was adopted to analyze the CS-based MUD problem.
Based on the analytic model, we derived the closed-form expression of the perfect AUD probability, the CE error, and the aggregate data rate of the CS-GF-NOMA mMTC system.
Then we optimized the EE and APCE of the GF-NOMA mMTC system via numerical method.
Simulation results verified the validity of our analysis and illustrated that CS-GF-NOMA had significantly improved AUD and data rate performances, compared with OP-GF-NOMA and GF-OMA.

In the future work, one direction is to study the secrecy rate of GF-NOMA because the low-rate short package transmission of IoT devices is with a high probability to be intercepted by an eavesdropper \cite{qi2020physical}.
Another direction is to futher develop a traffic-aware spatio-temporal model for CS-GF-NOMA with consideration of repetition slotted ALOHA protocal, which is widely adopted by the existing IoT applications \cite{yu2019analysis}.

\section*{Appendix}

\subsection{Proof of Lemma 1}

The proof of Lemma 1 is directly derived from \cite{wainwright2009sharp},
where \citep[Theorems 3 and 4]{wainwright2009sharp} respectively give the conditions of the achievability and the inachievability of LASSO-based noisy sparsity pattern recovery,
which are respectively the conditions to guarantee the perfect success of LASSO and to cause the failure of LASSO.

\subsubsection{Achievability}

From \citep[Theorem 3]{wainwright2009sharp}, if inequity
\begin{align}\label{lasso_cond1}
	\frac{c_2 M}{2K\log (N-K)}>1+\frac{\sigma^2}{\gamma^2K}
\end{align}
holds, then $\mathsf{Supp}(\widehat{\mathbf{q}}) \subset \mathsf{Supp}(\mathbf{q})$ with probability converging to one.
This indicates that the set of the detected MTCDs by LASSO is a subset of the set of the active MTCDs.
Furthermore, if $q_\text{min}>\upsilon $ also holds,
then LASSO can recover the exact sparsity profile of $\bm{q}$,
and the recovery error of each entry can be bounded as $\max_{n\in\mathsf{Supp(\mathbf{q})}} |q_n - \widehat{q}_n| \leq \upsilon$.

From (\ref{lasso_cond1}), we have
\begin{align}\label{lasso_cond1_1}
	K < \frac{c_2 M}{2\log (N-K)} - \frac{M}{2c_1 \log N}.
\end{align}

Since the MTCDs in the mMTC scenario are generally with very low activity,
$N\gg K$ and thus $N-K\approx N$.
Based on the above approximation, (\ref{lasso_cond1_1}) can be expressed as
\begin{align}
	K < \frac{M}{2 \log N}\left(c_2 -\frac1{c_1}\right).
\end{align}

\subsubsection{Inachievability}

On the other hand, from \citep[Theorem 4]{wainwright2009sharp}, if inequity
\begin{align}\label{lasso_fail_cond}
\frac{c_2 M}{2K\log (N-K)} < 1+\frac{\sigma^2}{\gamma^2K}
\end{align}
holds, then with probability converging to one,
LASSO cannot recovery the correct support of the original sparse vector.
With the approximation $N-K\approx N$, (\ref{lasso_fail_cond}) can be expressed as (\ref{K_max_fail}).

\subsection{Proof of Theorem 1}

In the SG network model, the number $K$ of active MTCDs is a Poisson random variable with parameter $\lambda$. Therefore, $\Pr\{K=k\} = e^{-\lambda}\lambda^k/k!$.
From Lemma 1, the probability of perfect LASSO AUD shown as (\ref{pr_succ_lasso1}) is evaluated as
\begin{align}\label{pr_succ_lasso}
  \mathcal{P}_\text{PER} = &\Pr \left\{ K <K_\text{max}, q_{\text{min}}>\upsilon \right\}   \nonumber\\
  = &\sum_{k=0}^{K_\text{max}} \Pr\{ K = k \} \Pr\{ q_{\text{min}}>\upsilon | K = k \}        \nonumber\\
  = & \sum_{k=0}^{K_\text{max}} \frac{e^{-\lambda}\lambda^k}{k!} \prod_{i=1}^{k}
      \Pr\left\{ \xi_{n_i}> \frac{\upsilon^2}{P} r_{n_i}^\alpha \right\}      \nonumber\\
  = & \sum_{k=0}^{K_\text{max}} \frac{e^{-\lambda}\lambda^k}{k!} \mathcal{P}_0^k.
\end{align}
where $\mathcal{P}_0$ is the probability that the received power of an active MTCD is greater than $\upsilon^2$, which can be evaluated as
\begin{align}\label{eq_P0}
  \mathcal{P}_0 = & 
      \mathbb{E}\left[ \Pr\left\{ \left. \xi>\frac{\upsilon^2}{P} r^\alpha \right| r \right\} \right] 
  = \mathbb{E}\left[ \left.e^{ - \frac{\upsilon^2}{P} r^\alpha } \right| r \right] \nonumber\\
  = & \int_{D_0}^{D_1} e^{ - \frac{\upsilon^2}{P} r^\alpha } f_{R}(r) \mathrm{d} r
  = \frac{2}{D_1^2-D_0^2} \underbrace{\int_{D_0}^{D_1} r e^{ -\frac{\upsilon^2}{P} r^\alpha} \mathrm{d} r}_{Q_1}.
\end{align}
Let $t=\frac{\upsilon^2}{P}r^{\alpha}$. Then $r=\left(\frac{P}{\upsilon^2}t\right)^{\frac1{\alpha}}$ and
$\mathrm{d}r = \frac1{\alpha} \left(\frac{P}{\upsilon^2}\right)^{\frac1{\alpha}} t^{\frac1{\alpha}-1} \mathrm{d}t$.
Integral $Q_1$ in (\ref{eq_P0}) can be evaluated as
\begin{align}\label{Q_1}
Q_1 = & \frac1{\alpha}\left(\frac{P}{\upsilon^2}\right)^{\frac{2}{\alpha}} 
			\int_{D_0}^{D_1} t^{\frac{2}{\alpha}-1}e^{-t}\mathrm{d}t \nonumber\\
= & \frac1{\alpha}\left(\frac{P}{\upsilon^2}\right)^{\frac{2}{\alpha}} 
	\left[ \Gamma\left(\frac{2}{\alpha},\frac{\upsilon^2 D_0^\alpha}{P} \right)-
			\Gamma\left(\frac{2}{\alpha},\frac{\upsilon^2 D_1^\alpha}{P} \right) \right] .
\end{align}
Substituting (\ref{Q_1}) into (\ref{eq_P0}), we can obtain (\ref{pr_succ_lasso_fin1}).
%
%

\subsection{Proof of Lemma 2}

From \citep[Theorem 1]{coluccia2014exact}, the LS estimation of $\mathbf{q}_{\mathcal{S}_0}$ based on observation $\mathbf{y} = \mathbf{\Phi}_{\mathcal{S}_0}\mathbf{q}_{\mathcal{S}_0} + \mathbf{\Phi}_{\mathcal{S}_1}\mathbf{q}_{\mathcal{S}_1} + \mathbf{w}$ yields
\begin{align}\label{mse1}
    \mathsf{MSE}_J  = \frac{\mathbb{E} \left[ \left\| \mathbf{\Phi}_{\mathcal{S}_1}\mathbf{q}_{\mathcal{S}_1} + \mathbf{w} \right\|^2 \right]}{M(M-J-1)}
    = \frac{\mathbb{E} \left[ \left\| \mathbf{\Phi}_{\mathcal{S}_1}\mathbf{q}_{\mathcal{S}_1} \right\|^2 \right] + M\sigma^2}{M(M-J-1)},
\end{align}
where $\mathbb{E} \left[ \left\| \mathbf{\Phi}_{\mathcal{S}_1}\mathbf{q}_{\mathcal{S}_1} \right\|^2 \right]
    = \mathbb{E} \Big[ \mathbf{q}_{\mathcal{S}_1}^H \mathbb{E} \left[ \mathbf{\Phi}_{\mathcal{S}_1}^H
        \mathbf{\Phi}_{\mathcal{S}_1} \right] \mathbf{q}_{\mathcal{S}_1}  \Big]
    = M \mathbb{E}\left[ \left\| \mathbf{q}_{\mathcal{S}_1} \right\|^2 \right]$.
Then (\ref{mse1}) can be expressed as
\begin{align}\label{mse2}
      \mathsf{MSE}_J = \frac{\mathbb{E}\left[ \| \mathbf{q}_{\mathcal{S}_1} \|^2 \right] + \sigma^2 }{M-J-1}.
\end{align}

We exploit Campbell's Theorem \cite{chiu2013stochastic} to derive $\mathbb{E}\left[ \| \mathbf{q}_{\mathcal{S}_1} \|^2 \right]$ in (\ref{mse2}).
Since $\left\| \mathbf{q}_{\mathcal{S}_1} \right\|^2 = \sum_{n\in\Phi}
 \mathbf{1}\left( P \xi_n r_n^{-\alpha}<\upsilon^2 \right) \xi_n r_n^{-\alpha}$,
where $\mathbf{1}(\cdot)$ is the indicator function,
we have
\begin{align}\label{mse_eq}
	& \mathbb{E}\left[ \| \mathbf{q}_{\mathcal{S}_1} \|^2 \right]
	= \frac{\lambda P}{\pi(D_1^2-D_0^2)} \int_{\mathbb{R}^2} r^{-\alpha} \mathbb{E} \left[ 
			\mathbf{1}\left(\xi <\frac{\upsilon^2}{P} r^{\alpha} \right) \xi \right] \mathrm{d}x \notag\\
	&~~ = \frac{\lambda P}{\pi (D_1^2 - D_0^2) } \int_{D_0}^{D_1} r^{-\alpha} \int_{0}^{\frac{\upsilon^2}{P} r^{\alpha}}
			   \xi f_\xi(\xi) \mathrm{d}\xi \times 2\pi r \mathrm{d}r \notag\\
	&~~ = \frac{2 \lambda P}{D_1^2 - D_0^2} \int_{0}^{D_1} r^{1-\alpha} \mathrm{d}r
			\int_{0}^{\frac{\upsilon^2}{P} r^{\alpha}} \xi e^{-\xi} \mathrm{d}\xi \notag\\
	&~~ = \frac{2 \lambda P}{D_1^2 - D_0^2} \int_{0}^{D_1} r^{1-\alpha} \left[  
				1-e^{-\frac{\upsilon^2}{P} r^{\alpha}} - \frac{\upsilon^2}{P} r^{\alpha} e^{-\frac{\upsilon^2}{P} r^{\alpha}}
			\right]\mathrm{d}r \notag\\
	&~~ = \frac{2 \lambda P}{D_1^2 - D_0^2} \Bigg\lbrace \frac{D_0^{2-\alpha} - D_1^{2-\alpha}}{\alpha-2}
			- \frac1{\alpha} \left( \frac{P}{\upsilon^2}\right)^{\frac{2}{\alpha}-1} \notag \\
			&~~~~~ \times \Bigg[
				\Gamma\left( \frac{2}{\alpha} -1, \frac{\upsilon^2 D_0^\alpha}{P} \right) 
				- \Gamma\left( \frac{2}{\alpha} -1, \frac{\upsilon^2 D_1^\alpha}{P} \right) \notag \\
			&~~~~ + \Gamma\left( \frac{2}{\alpha}, \frac{\upsilon^2 D_0^\alpha}{P} \right) 
			- \Gamma\left( \frac{2}{\alpha}, \frac{\upsilon^2 D_1^\alpha}{P} \right) \Bigg] \Bigg\rbrace.
\end{align}
Substituting (\ref{mse_eq}) into (\ref{mse2}), we can obtain (\ref{eq_mse_thr}).

\subsection{Proof of Theorem 2}

The average NMSE of CE can be evaluated as
\begin{align}\label{anmse}
	\overline{\mathsf{NMSE}} = \left.\sum\nolimits_{j=1}^{K_\mathrm{max}} \Pr\{J=j\} \mathsf{MSE}_j
	\right/\Xi,
\end{align}
where the probability $\Pr\{J=j\}$ that $j$ active MTCDs are detected by LASSO AUD can be evaluated as
\begin{align}\label{Pr_j}
	\Pr\{J=j\} = & \sum_{k=j}^{K_\mathrm{max}} \Pr\{K=k\} \times \Pr\{J=j|K=k\} \notag\\
	 = &\sum_{k=j}^{K_\mathrm{max}} \frac{e^{-\lambda}\lambda^k}{k!} \times \tbinom{k}{j} \mathcal{P}_0^j (1-\mathcal{P}_0)^{k-j},
\end{align}
and the expectation $\Xi=\mathbb{E}\left[|{q}_{n}|^2\right]$ of the AP received power of a detected MTCD $n\in\mathcal{S}_0$ is
\begin{align}\label{E_q0}
\Xi = & \int_{D_0}^{D_1} \int_{\frac{\upsilon^2} {P} r^{\alpha}}^\infty P \xi r^{-\alpha}
    \frac{f_\xi(\xi) f_r(r)}{\mathcal{P}_0} \mathrm{d}\xi \mathrm{d}r \notag\\
= & \frac{2 P}{\mathcal{P}_0(D_1^2 - D_0^2)} \int_{D_0}^{D_1}  \left[  
 		r^{1-\alpha} e^{-\frac{\upsilon^2}{P} r^{\alpha}} + \frac{\upsilon^2}{P} r e^{-\frac{\upsilon^2}{P} r^{\alpha}}
		\right]\mathrm{d}r \notag\\
= & \frac{2P^{\frac{2}{\alpha}}}{\alpha \mathcal{P}_0 (D_1^2 - D_0^2) {\upsilon}^{\frac{4}{\alpha}-2}} \notag\\
& \times \Bigg[ \Gamma\left( \frac{2}{\alpha} -1, \frac{\upsilon^2 D_0^\alpha}{P} \right) 
	- \Gamma\left( \frac{2}{\alpha} -1, \frac{\upsilon^2 D_1^\alpha}{P} \right) \notag \\
& + \Gamma\left( \frac{2}{\alpha}, \frac{\upsilon^2 D_0^\alpha}{P} \right) 
	- \Gamma\left( \frac{2}{\alpha}, \frac{\upsilon^2 D_1^\alpha}{P} \right) \Bigg] .
\end{align}

Substituting (\ref{Pr_j}) and (\ref{E_q0}) into (\ref{anmse}), we can obtain (\ref{th_2_nmse}).

\subsection{Proof of Theorem \ref{Th_average_rate}} \label{proof_Th_average_rate}

The average aggregate rate can be evaluated as
\begin{align}\label{ave_rsum}
	& \overline{R}_\mathrm{A} = \mathbb{E}\left[ \left.
		\log_2\left( \sum_{k=1}^{K}\left| {q}_{n_k} \right|^2 + M \sigma^2 \right)\right|K\right] \notag\\
	&~~~~~~~~ - \mathbb{E}\left[
		\log_2\left(\sum\nolimits_{\left| q_n \right|<\upsilon} \left| q_n \right|^2 + M \sigma^2 \right)\right] \notag\\
	\geq & \sum_{k=1}^{K_{max}} \frac{e^{-\lambda}\lambda^k}{k!} \underbrace{ \mathbb{E}\left[ 
		\log_2\left( 1 + \sum_{k'=1}^{k}\frac{\left|{q}_{n_{k'}}\right|^2}{M \sigma^2} \right)\right]}_{Q_2}   + \log_2 \left(M \sigma^2 \right)\notag\\
	& -  \log_2\left( \mathbb{E}\left[\sum\nolimits_{\left| q_n \right|<\upsilon} \left|q_n\right|^2\right] + M \sigma^2  \right), 
\end{align}
where the inequality is according to Jensen's inequality \citep[Section 12.411]{izrail2007table} and the concavity of logarithmic functions.
Campbell's Theorem can be used to evaluate $\mathbb{E}\left[\sum\nolimits_{\left| q_n \right|<\upsilon} \left|q_n\right|^2\right]$ in (\ref{ave_rsum}), similar with (\ref{mse_eq}).
From \citep[Lemma 1]{hamdi2010useful}, $Q_2$ in (\ref{ave_rsum}) can be evaluated as
\begin{align}\label{average_rate}
	Q_2 = \mathbb{E}\left[ \log_2(1 + \Psi_k)\right] 
		= \frac1{\ln 2} \int_{0}^{\infty} \frac{e^{-s}}{s} \big(1-\mathcal{L}_{\Psi_k}(s)\big)\mathrm{d}s,
\end{align}
where $\Psi_k = \sum_{k'=1}^{k} \psi_{k'} 
	= \sum_{{k'}=1}^{K} \frac{\left| q_{n_{k'}}\right|^2}{M\sigma^2} $,
and $\mathcal{L}_{\Psi_K}(\cdot)$ denote the Laplace transform,
which can be evaluated as
\begin{align} \label{LapTrans}
\mathcal{L}_{\Psi_k}(s)= &\mathbb{E}_{\Psi_k}\left[e^ {-s\sum_{{k'}=1}^{k} \psi_{k'} } \right]
	= \mathbb{E}_{\Psi_k} \left[
		\prod\nolimits_{{k'}=1}^{k} e^{-s\psi_{k'} } \right]\notag \\
	= &\Big (\mathbb{E}_{ \psi_{k'} } 
		\Big[e^{-s \psi_{k'} } \Big] \Big)^{k}
	= \big( \mathcal{L}_{ \psi_{k'} }(s) \big)^k.
\end{align}

The CDF of $\psi_{k'}$ is evaluated as
\begin{align}\label{cdf_h}
F_Z(z) = & 1 - \Pr\left\{\frac{P \xi r^{-\alpha}}{M\sigma^2} > z\right\}= 1 - \mathbb{E}\left[
		 \xi > \frac{M\sigma^2 r^{\alpha}}{P} z \Big| r  \right] \notag\\
	= & 1- \frac{2}{D_1^2 - D_0^2} \int_{D_0}^{D_1} r e^{-\frac{M\sigma^2 r^{\alpha}}{P}z} \mathrm{d}r,
\end{align}
and then the PDF of $\psi_{k'}$ is calculated by the derivative of $F_Z(z)$ in (\ref{cdf_h})
\begin{align}\label{pdf_h}
f_Z(z) = \frac{2 M \sigma^2}{(D_1^2 - D_0^2)P} \int_{D_0}^{D_1} r^{\alpha+1} e^{-\frac{M\sigma^2}{P} r^{\alpha}z} \mathrm{d}r
\end{align}

Therefore, $\mathcal{L}_{\psi_{k'}}(s)$ in (\ref{LapTrans}) can be evaluated as
\begin{align}\label{LapTrans_single}
&\mathcal{L}_{\psi_{k'}}(s) = \int_{0}^{\infty} e^{-sz}f_Z(z) \mathrm{d}z \notag\\
	&= \frac{2 M\sigma^2}{(D_1^2 - D_0^2)P} \int_{0}^{\infty} e^{-sz} \mathrm{d}z
		\int_{D_0}^{D_1} r^{\alpha+1} e^{-\frac{M\sigma^2}{P} r^{\alpha}z} \mathrm{d}r \notag\\
	&= \frac{2 M\sigma^2}{(D_1^2 - D_0^2)P} \int_{D_0}^{D_1} r^{\alpha+1} \mathrm{d}r
		\int_{0}^{\infty} e^{-\left(s+\frac{M\sigma^2}{P} r^{\alpha}\right)z}   \mathrm{d}z   \notag\\
	&= \frac{2 M\sigma^2}{(D_1^2 - D_0^2)P} \int_{D_0}^{D_1}
		\frac{r^{\alpha+1}}{s+\frac{M\sigma^2}{P}r^{\alpha}} \mathrm{d}r    \notag\\
	&= \frac{2 M\sigma^2}{(D_1^2 - D_0^2)sP} \notag\\ 
	&~~~ \times\underbrace{ \left[\int_{0}^{D_1}
		\frac{r^{\alpha+1}}{1+\frac{M\sigma^2}{sP}r^{\alpha}} \mathrm{d}r \right.}_{Q_3} 
		- \underbrace{ \left.\int_{0}^{D_0}
			\frac{r^{\alpha+1}}{1+\frac{M\sigma^2}{sP}r^{\alpha}} \mathrm{d}r \right]}_{Q_4} 
\end{align}
Let $t = (r/D_1)^\alpha$. Then $\mathrm{d}r = \frac1{\alpha}D_1 t^{\frac1{\alpha}-1}\mathrm{d}t$, and $Q_3$ can be evaluated as
\begin{align}\label{Q_3}
	Q_3 = & \frac{D_1^{\alpha+2}}{\alpha}
		\int_{0}^{1} \frac{t^{\frac{2}{\alpha}}}{1+\frac{M\sigma^2}{sP}D_1^\alpha t} D_1^\alpha \mathrm{d}t \notag\\
		= & \frac{D_1^{\alpha+2}}{\alpha B\left(\frac{2}{\alpha}+1,1\right)}
		F\left( 1, \frac{2}{\alpha}+1; \frac{2}{\alpha}+2;-\frac{M\sigma^2}{sP} D_1^\alpha \right),
\end{align}
where $B\left(z,w\right) = \int_{0}^{1}t^z (1-t)^w\mathrm{d}t$ is the Beta function. 
$B\left(\frac{2}{\alpha}+1,1\right) = \frac{2}{\alpha}+1$.
Similarly, $Q_4$ can be evaluated as
\begin{align}\label{Q_4}
	Q_4 = \frac{D_0^{\alpha+2}}{\alpha B\left(\frac{2}{\alpha}+1,1\right)} 
		F\left( 1, \frac{2}{\alpha}+1; \frac{2}{\alpha}+2;-\frac{M\sigma^2}{sP} D_0^\alpha \right).
\end{align}

Finally, we can obtain (\ref{ave_R}) by combining
(\ref{ave_rsum}), (\ref{average_rate}), (\ref{LapTrans}), (\ref{LapTrans_single}), (\ref{Q_3}), and (\ref{Q_4}).

\bibliographystyle{ieeetr}
\footnotesize
\bibliography{references}
\end{document}